\RequirePackage{fix-cm}
\documentclass[twocolumn]{svjour3}          
\smartqed  
\usepackage{graphicx}
\usepackage{amsmath}
\usepackage{url}
\usepackage{float}
\usepackage{hyperref}
\usepackage[misc]{ifsym}
\begin{document}

\title{HoloLens 2 Technical Evaluation as Mixed Reality Guide 
}

\titlerunning{HoloLens 2 Technical Evaluation as Mixed Reality Guide}        

\author{Hung-Jui~Guo
        \href{https://orcid.org/0000-0003-2233-846X}{\includegraphics[scale=0.02]{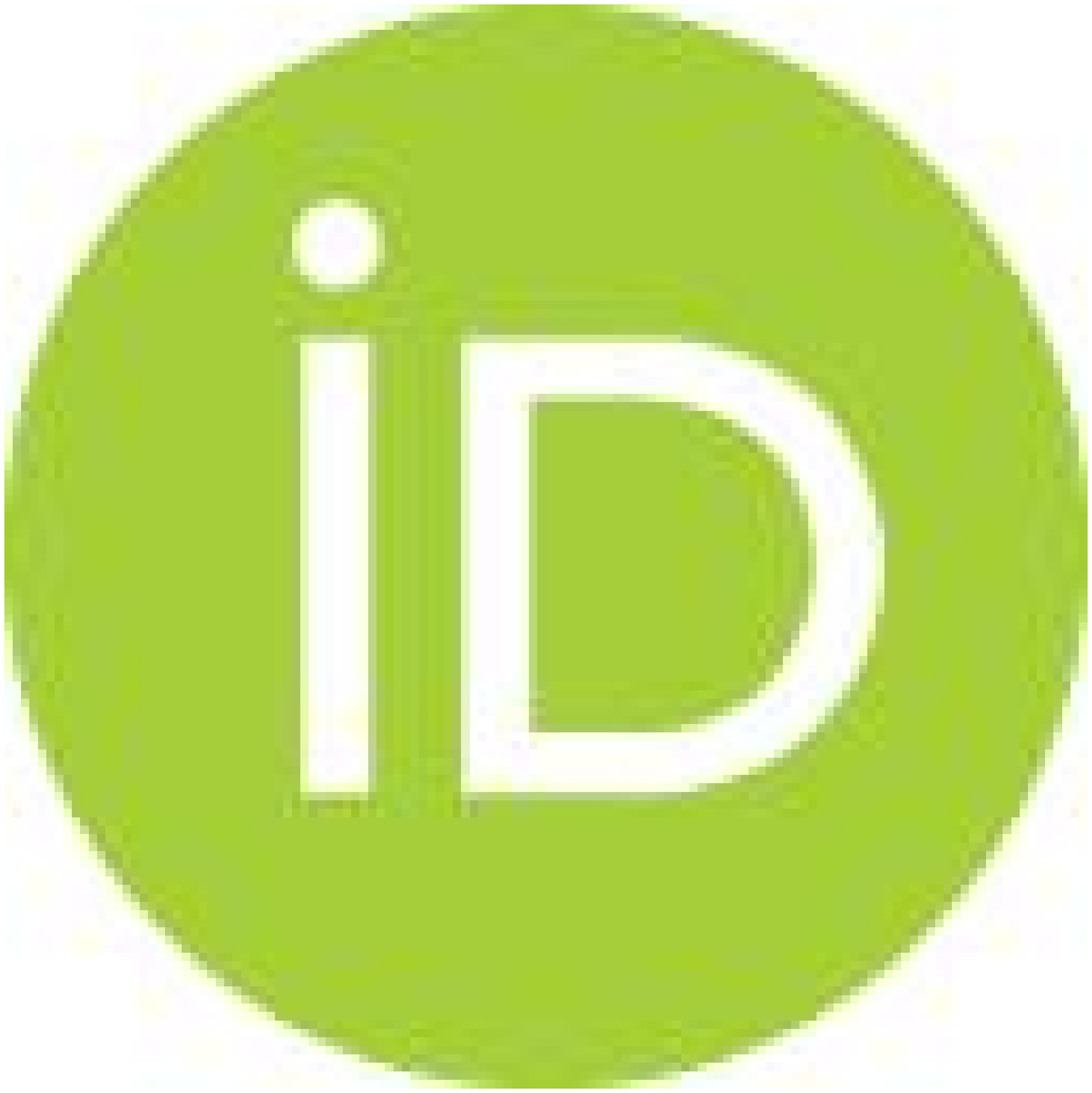}}%
        \and
        Balakrishnan~Prabhakaran
        \href{https://orcid.org/0000-0003-0385-8662}{\includegraphics[scale=0.02]{ORCIDiD_icon128x128.pdf}} 
}

\authorrunning{H.J., Guo, B. Prabhakaran} 

\institute{\textrm{\Letter} Hung-Jui~Guo \at
              \email{hxg190003@utdallas.edu}             \\
            Software Engineering Department, the Univerity of Texas at Dallas, Texas, United State
           \and
            Balakrishnan~Prabhakaran 
            \at 
              \email{bprabhakaran@utdallas.edu} \\
              Computer Science Department, the Univerity of Texas at Dallas, Texas, United State
}

\date{Received: date / Accepted: date}

\maketitle
\begin{abstract}
Mixed Reality (MR) is an evolving technology lying in the continuum spanned by related technologies such as Virtual Reality (VR) and Augmented Reality (AR), and creates an exciting way of interacting with people and the environment. This technology is fast becoming a tool used by many people, potentially improving living environments and work efficiency. Microsoft HoloLens \cite{hololensWeb} has played an important role in the progress of MR, from the first generation to the second generation. In this paper, we systematically evaluate the functions of applicable functions in HoloLens 2 \cite{ungureanu2020hololens}. These evaluations can serve as a performance benchmark that can help people who need to use this instrument for research or applications in the future. The detailed tests and the performance evaluation of the different functionalities show the usability and possible limitations of each function. We mainly divide the experiment into the existing functions of the HoloLens 1, the new functions of the HoloLens 2, and the use of research mode. This research results will be useful for MR researchers who want to use HoloLens 2 as a research tool to design their own MR applications.
\keywords{Mixed Reality \and HoloLens \and HoloLens 2 \and hologram \and Inertial Measurement Unit (IMU) \and Spatial Anchor}
\end{abstract}

\section{Introduction}
Mixed Reality (MR) \cite{mixedRealityOriginal}, \cite{Milgram1995} has introduced an exciting new way of interaction among people as well as between human and the real-world. MR can play a very important role in the current demand for remote work and remote communication. MR is being applied in many different fields, such as medical science, education or industrial design. Many new research works are being carried out to integrate MR into daily life activities of the general public, thereby increasing convenience and work efficiency. \par
\begin{figure}[htbp]
\centering
\includegraphics[width=8.4cm, keepaspectratio=true]{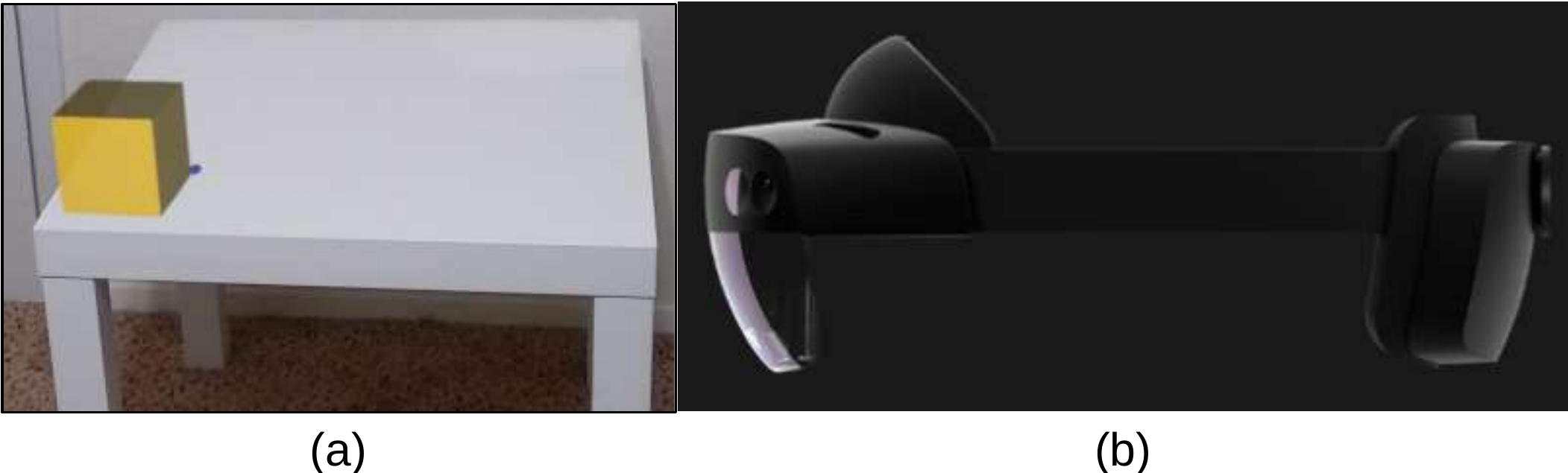}
\caption{(a) An example of putting virtual object on table in the real environment. (b) HoloLens 2 device (Picture taken from Microsoft HoloLens 2 webcite \cite{hololensWeb})}
\label{fig:Intro}
\vspace{-1em}
\end{figure}

Mixed Reality (MR) is in the continuum spanned by related technologies such as Virtual Reality (VR) \cite{burdea2003virtual} and Augmented Reality (AR) \cite{azuma1997survey}. VR changes the entire surrounding environment into a virtual form and allow interactions between the user and the virtual word. AR augments the real-world by including graphics, sounds, and touches feedback through AR devices. However, AR does not facilitate interactions between users and the virtual objects added to the real-world scene. MR combines the advantages of the two, scanning the surrounding real-world environment to build a model, and then adds the needed virtual object in this environment. Users can directly interact with virtual objects (by operations such as scaling, rotation, or translation) in the real environment using their hands (or other devices). Furthermore, virtual objects can also interact with real objects in the mixed reality world. For example, a virtual cube can be placed on a table in the real-world, as shown in Figure \ref{fig:Intro}a. \par

Microsoft HoloLens 2 \cite{hololensWeb} (Figure \ref{fig:Intro}b) is an updated version of the previous generation HoloLens 1 headset from Microsoft. Compared with the previous generation, the HoloLens 2 has improved most of the features, such as display resolution, field of view, weight and so on. Also, some new functions and sensors have been added, such as eye-tracking system, biometric security system, two hands tracking, and Inertial Measurement Unit (IMU) sensors. \par
\begin{figure}[htbp]
\centering
\includegraphics[width=8.4cm, keepaspectratio=true]{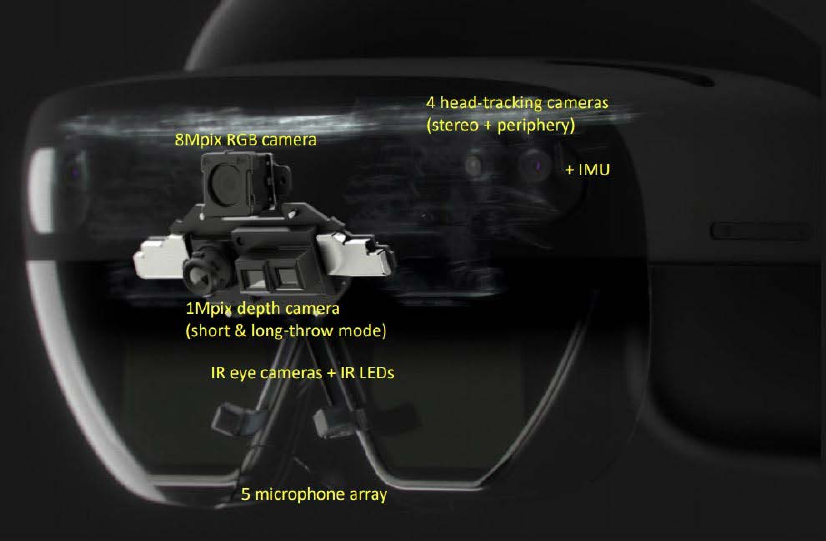}
\caption{HoloLens 2 input sensors \cite{ungureanu2020hololens}}
\label{fig:Sensors}
\vspace{-1em}
\end{figure}

In addition to the hardware updates, the corresponding software updates are also very impressive, especially the newer version of research mode \cite{ungureanu2020hololens}. Users can access multiple sensors by using a set of C++ APIs and tools of the research mode in HoloLens 2. The sensors available for access include Visible Light cameras, IR (Infra-Red), Depth camera, and IMU sensors, as shown in Figure \ref{fig:Sensors}. These will be of significant help to researchers who need to develop sensor-based applications in the future. \par
In this paper, we introduce a series of performance evaluation tests on most available sensors and features in HoloLens 2. Part of the tests in this paper refers to a HoloLens 1 paper \cite{Liu2018} to test the performance improvements, as compared between HoloLens 2 and HoloLens 1. These experimental results can help us understand the capabilities and possible limitations of HoloLens 2. We also test and discuss the performance of functions in the research mode. \par
This paper is organized as follows. Section II discusses the works related to mixed reality and HoloLens. Section III introduces the experimental design we used to test the device. Then Section IV shows the detailed experiment procedure according to the previous section, and Section V provides the results and discussions on each experiment. Lastly, Section VI concludes the whole paper and sums up our contributions.


\section{Related work}

\subsection{Mixed Reality}
\cite{Dey18} describes several AR-related user studies and provide ideas for deriving MR experiments from AR experiments. In \cite{Christopher2021}, the authors unified the AR applications and provided a link between testing and implementation, which can be leveraged for MR applications as well. \par
In order to unify the cognitive differences arising from the development of the MR field in recent years, \cite{speicher2019mixed} conducted interviews with a number of experts and referred to various documents to identify the concepts related to MR. \cite{benford1998understanding} proposed methods for applying it to stage a poetry performance simultaneously within real and virtual theaters. \cite{Billinghurst1999} proposed a collaborative MR interface, which is a direction that many other researchers have focused on in recent years. In addition, some other researchers used this technology in aesthetic design \cite{fiorentino2002spacedesign}, education \cite{Hughes2005}, \cite{Pan2006}, \cite{Tang2020}, and entertainment \cite{Cheok2009}, \cite{Coutrix2006}. In \cite{carotenuto2018indoor}, the authors used MR technology to implement an indoor positioning system. Since location information is also an important part of MR research, this paper will be a benchmark for future related research. Lastly, in \cite{costanza2009mixed} and \cite{Bekele2018}, the authors made a detailed survey on the field of MR to help other researchers increase their understanding of this field.

\subsection{HoloLens 1}
The first generation of HoloLens, HoloLens 1, \cite{hololensWeb} was first released in 2016 by Microsoft. HoloLens 1 is a pair of mixed reality smart-glasses, which is the first head-mounted display device running the Windows Mixed Reality platform. Since this device has been available for several years, there have been many research topics and applications related to it. \par
In paper \cite{Chen2015}, the authors proposed a method that uses only one HoloLens 1 and several Skype-enabled devices such as tablets or PCs to achieve remote cooperation. In addition to the aforementioned applications, HoloLens 1 can be used in the medical field. For example, \cite{Aruanno2017} tried to use this device as a therapeutic tool for people with Alzheimer's Disease, and \cite{Hanna2018} used HoloLens 1 in anatomic pathology to test clinical and non-clinical applications. Moreover, \cite{Wangm2018} proposed a study of mixing HoloLens 1 and 3D geographic information, and \cite{Garon2017} provided a method to add higher resolution depth data in the device. \cite{Evans2017} and \cite{Liu2018} implemented a series of evaluation tests to evaluate the performance, advantages, and disadvantages of HoloLens 1. The detailed test contents of paper \cite{Liu2018} are the main reference and comparison article in this paper; therefore, we use many similar tests in our experiments for the purpose of comparison.

\subsection{HoloLens 2}
It has only been about two and a half years since the release of HoloLens 2 \cite{hololensWeb}, so the number of related papers is relatively small. According to paper \cite{OConnor2019}, the authors introduced newly added functions, capability, and HPU upgrades in HoloLens 2. Then, \cite{ungureanu2020hololens} announced a new version of research mode for HoloLens 2, allowing users to use this tool to access individual sensors in the device. This function could help many researchers to obtain more sensorial data according to different needs. \cite{vidal2020analysis} applied HoloLens 2 in the field of education to ease the interaction between teachers and students, and \cite{nikolov2020visualization} used the device for industrial purposes. \cite{Park2020} used HoloLens 2 to increase the success rate of medical operations and reduce the operation time. Due to the COVID-19 situation, \cite{Martin2020} and \cite{levy2021mixed} proposed using HoloLens 2 for remote care to reduce the contact between doctors and patients. Currently, there is still no comprehensive set of performance evaluation, testing the available features of HoloLens 2 in detail. Therefore, we carried out various experiments testing the different functionalities of Hololens 2, and the results reported in this paper can serve both as a benchmark and as design criteria for MR applications.


\section{Experiment Design and Procedure}
To evaluate the functionality of HoloLens 2, we design several different experiments to examine its features. All testing apps are created in Unity 3D or using C++ APIs and exported to the HoloLens 2 device with software version 20H2. We also provide the experiment procedure detail about how the testers need to do each test. The same environmental condition names in each section have the same parameters.

\subsection{Hand Tracking}
By using the AHAT (Articulated HAnd Tracking) depth camera \cite{ungureanu2020hololens}, the HoloLens 2 can capture hand movements to obtain hand tracking data. We can access this data by using either research mode or Unity 3D apps, and the output results are the same. Therefore, for convenience of testing, we use Unity 3D apps as our main testing tool in this evaluation process. \par
\begin{figure}[htbp]
\centering
\includegraphics[width=8.4cm, keepaspectratio=true]{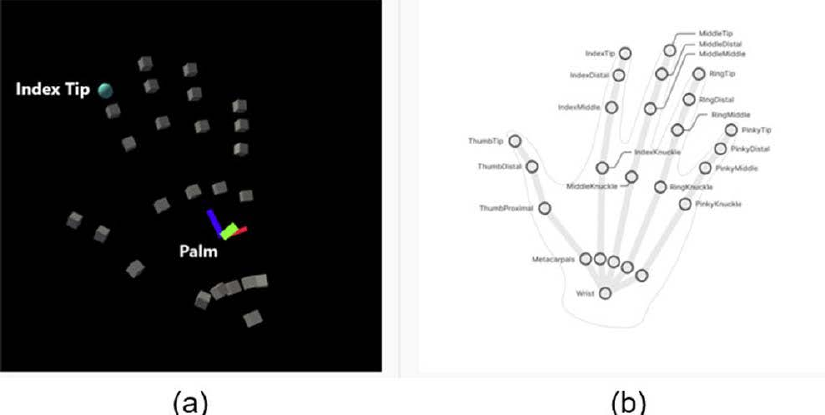}
\caption{Examples of hand joint feature (a) Default hand joint hologram representation in HoloLens 2 world (b) Real-world hand joint labels (Picture taken from Microsoft HoloLens 2 webcite \cite{hololensWeb})}
\label{fig:joint}
\vspace{-1em}
\end{figure}

This experiment evaluates the hand tracking feature in HoloLens 2 by calculating the difference between hand joints hologram and real-world human hand joints. Figure \ref{fig:joint}a shows the default hand joint hologram generated in HoloLens 2 world, and Figure \ref{fig:joint}b shows the corresponding joint name of the human hand. \par
\begin{figure}[htbp]
\centering
\includegraphics[width=8.4cm, keepaspectratio=true]{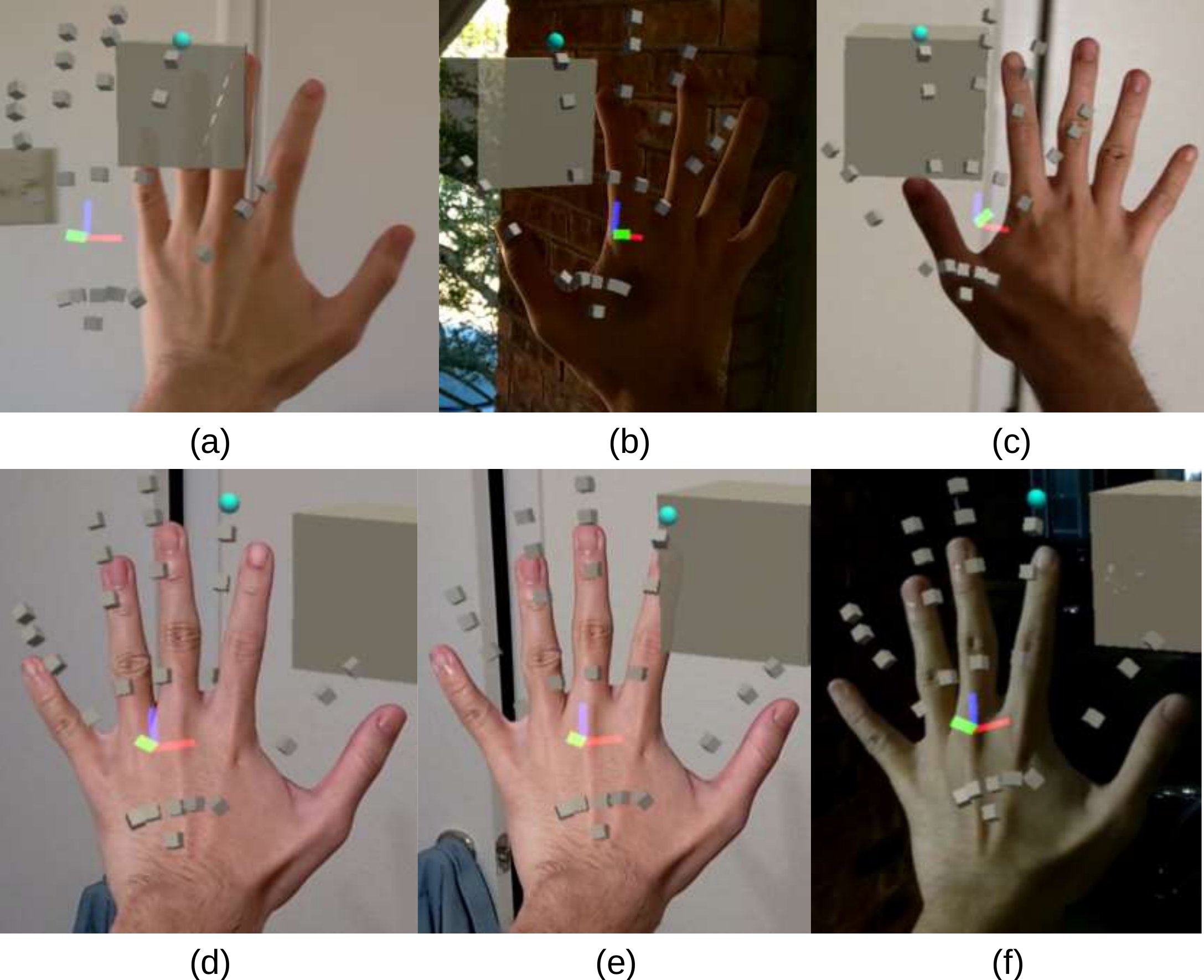}
\caption{Steady hand tracking result under different environmental conditions (a) Daylight indoor (b) Daylight outdoor (c) Night time with lamp (d) Night time indoor with headlight and lamp (e) Night time indoor with headlight (f) Night time outdoor with lamp}
\label{fig:hand}
\vspace{-1em}
\end{figure}

Since users implement apps in many different conditions, our experiments cover as many environmental factors as possible to simulate different situations. In this case, we choose to examine the hand tracking feature of HoloLens 2 under daylight/night time and with/without headlight/lamp. In addition to different environmental variables, we also test this feature at different hand moving speeds, such as 1/0.5/0.2 meters per second. The hand tracking performance is evaluated by calculating the average Euclidean distances between each hand joint point in the hologram and each real-world human hand joint point.
When projecting hand joint feature from HoloLens 2 world space to 2D image space, the projection will produce a slight error between hologram hand and real-world hand, as shown in Figure \ref{fig:hand}. Hence we also test the hand tracking feature when testers stop moving their hands as a baseline to calculate the true error.

To perform the hand tracking test, we first create a virtual box (0.1 m X 0.1 m X 0.1 m) in Unity 3D and make it move left and right like a pendulum at a fixed speed,  0.2 m / second, 0.5 m / second, and 1 m / second with moving distance 1 m. Then we ask the testers to open their hands and wave either their left or right hand back and forth at the moving cube's speed to record the generated hand joints as the result, as shown in Figure \ref{fig:joint}. Besides, we also record the hand joints result when testers' hands stop moving. Apart from the hand moving speed, we also use different environmental variables (daylight indoor, daylight outdoor, nighttime indoor with head light and lamp, nighttime indoor with head light, nighttime indoor with lamp, and nighttime outdoor with lamp) as one of our test conditions. Afterward, we record testers' hand waving hand joints tracking video in different conditions and take out five frames as test results.

\subsection{Eye Tracking}
HoloLens 2 Eye Tracking API provides a single eye-gaze ray for developers to access eye-tracking data for further use in eye-tracking experiments. This feature is newly introduced, compared to the previous generation HoloLens 1 \cite{Liu2018}. In this experiment, we test the eye-tracking functionality under different environmental conditions. At the same time, we evaluate the eye-tracking feature on real-world objects and virtual world objects on top of these conditions. During the eye-tracking procedure, we will ask the testers to gaze at each target point for two seconds according to our instructions and take pictures three times of the results for average purposes. Then, we use the captured pictures to calculate the error distance. The eye-tracking performance is then evaluated in terms of Euclidean distances between the eye-gaze ray and target points.
\begin{enumerate}
\item Virtual Objects Eye Gaze Tracking: \hspace{\fill}\\
In the virtual objects eye gaze tracking, we create five red virtual cubes in five different directions with size 1 cm X 1 cm X 1 cm, and we also generate a black platform behind these five boxes to visualize the eye gaze point on the same flat surface, as shown in Figure \ref{fig:eye1}. Then we ask the testers to gaze five virtual cubes under different environmental conditions, daylight/night time an with/without headlight/lamp. We use the same environmental variables of the hand tracking experiment. When the testers arrive at the selected location, we ask them to start the app and gaze at each point for two seconds; then, we record the test process and extract the result accordingly.

\begin{figure}[htbp]
\centering
\includegraphics[width=8.4cm, keepaspectratio=true]{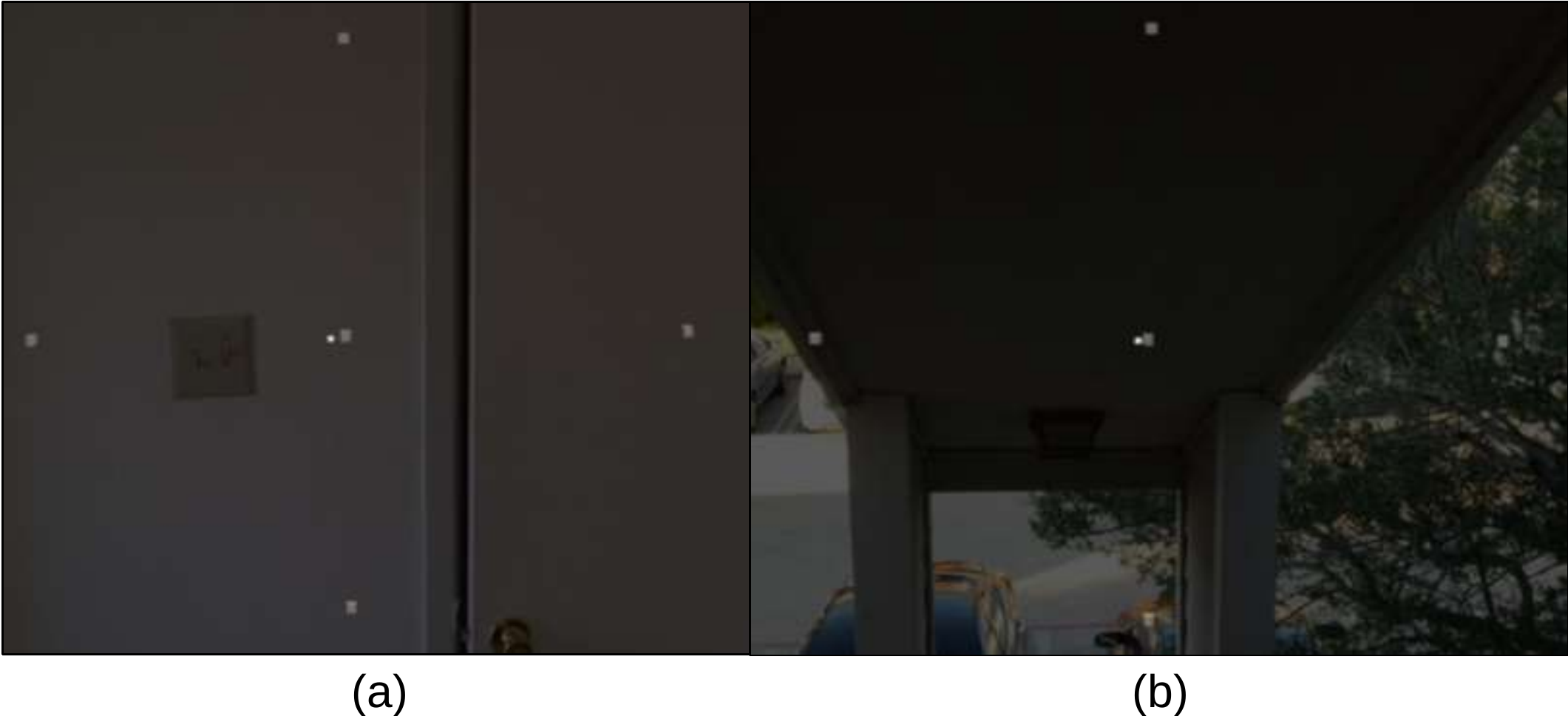} 
\caption{Eye-gaze tracking examples, the white dot represents the tester's eye-gaze point, and gray boxes indicate the target virtual cubs (a) Daylight indoor (b) Daylight outdoor}
\label{fig:eye1}
\vspace{-1em}
\end{figure}

\item Real-world Objects Eye Gaze Tracking: \hspace{\fill}\\
Real-world objects eye gaze tracking was performed to evaluate the eye-tracking accuracy in a real-world environment. To complete this experiment, we put five different real-world black dots on a box and ask the testers to gaze at these five points, as shown in Figure \ref{fig:eye2}. We only choose two environmental variables - Daylight indoor and nighttime with lamp - as our control variable since the target is difficult to see clearly under the environment with insufficient lighting. We then ask the testers to walk to the selected location, start the app, and gaze at each black dot for two seconds; then, we record the test process and take a screenshot from the recorded video to extract the evaluation result accordingly.

\begin{figure}[htbp]
\centering
\includegraphics[width=8.4cm, keepaspectratio=true]{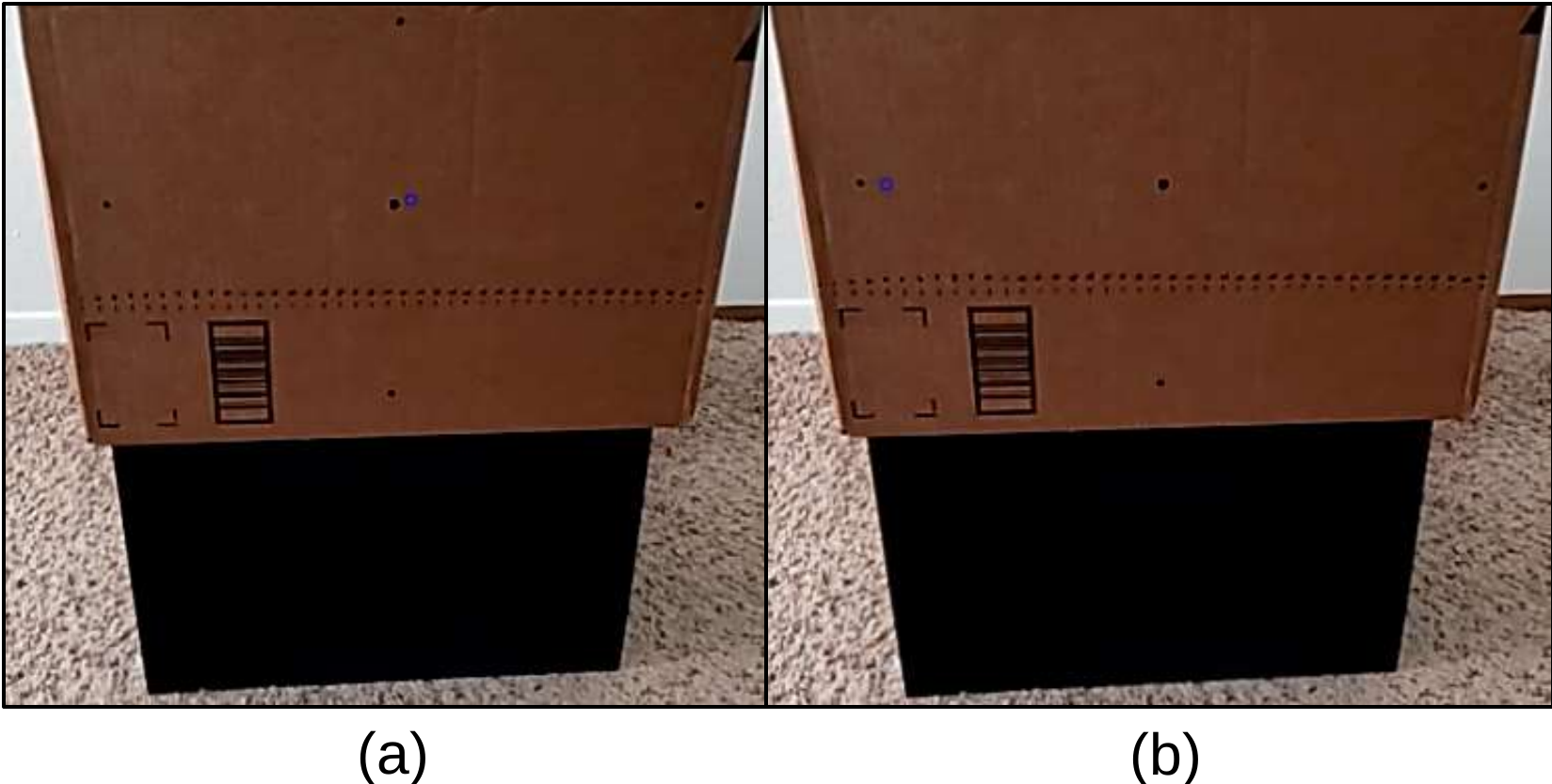}
\caption{Real-world eye gaze tracking examples under daylight, the blue dot represents the tester's eye-gaze point, and black dots indicate the target virtual cubes}
\label{fig:eye2}
\vspace{-1em}
\end{figure}

\end{enumerate}

\subsection{Spatial Mapping}
Spatial mapping is the key in the hologram experiment, which uses the long-throw depth camera. We can access spatial mapping data through either research mode or Unity 3D apps, and the results are the same. For testing purposes, we use Unity 3D apps as our main testing platform. This spatial mapping experiment evaluates the reality of holograms, which is generated by HoloLens 2. According to the previous HoloLens 1 paper \cite{Liu2018}, the authors assess the result by overlapping the holograms on the target real-world objects. Therefore, we use the same concept to design our experiments and use the same accuracy deviation to evaluate the result in Hololens 2. To complete this experimental design, we create a HoloLens 2 app with Unity 3D to make spatial mapping visible. Whenever we need to reconstruct a new hologram, we ask the testers delete all holograms in HoloLens 2 via menu setting. 
\begin{enumerate}
\item Hologram Visualization: \hspace{\fill}\\
In the hologram visualization experiment, we use a HoloLens 2 product box as a real-world standard to create a hologram box in the mix-reality world. In this case study, this box's size is 37.8 cm X 24.7 cm X 26.8 cm. Then we create a red hologram box of the same size for overlapping, which is shown in Figure \ref{fig:hologram}. Besides, we constrain the hologram box so that it cannot rotate or change size in order to avoid the testers changing the shape. During this process, we ask the testers to stand in front of the real-world box and start the testing app. The testers are asked to move the visualized hologram box to overlap with the real-world box as exactly as possible. Then we will take pictures from the real-world box's front, side and top views, as shown in Figure \ref{fig:hologram}. These pictures will be used as our evaluation result.

\begin{figure}[htbp]
\centering
\includegraphics[width=8.4cm, keepaspectratio=true]{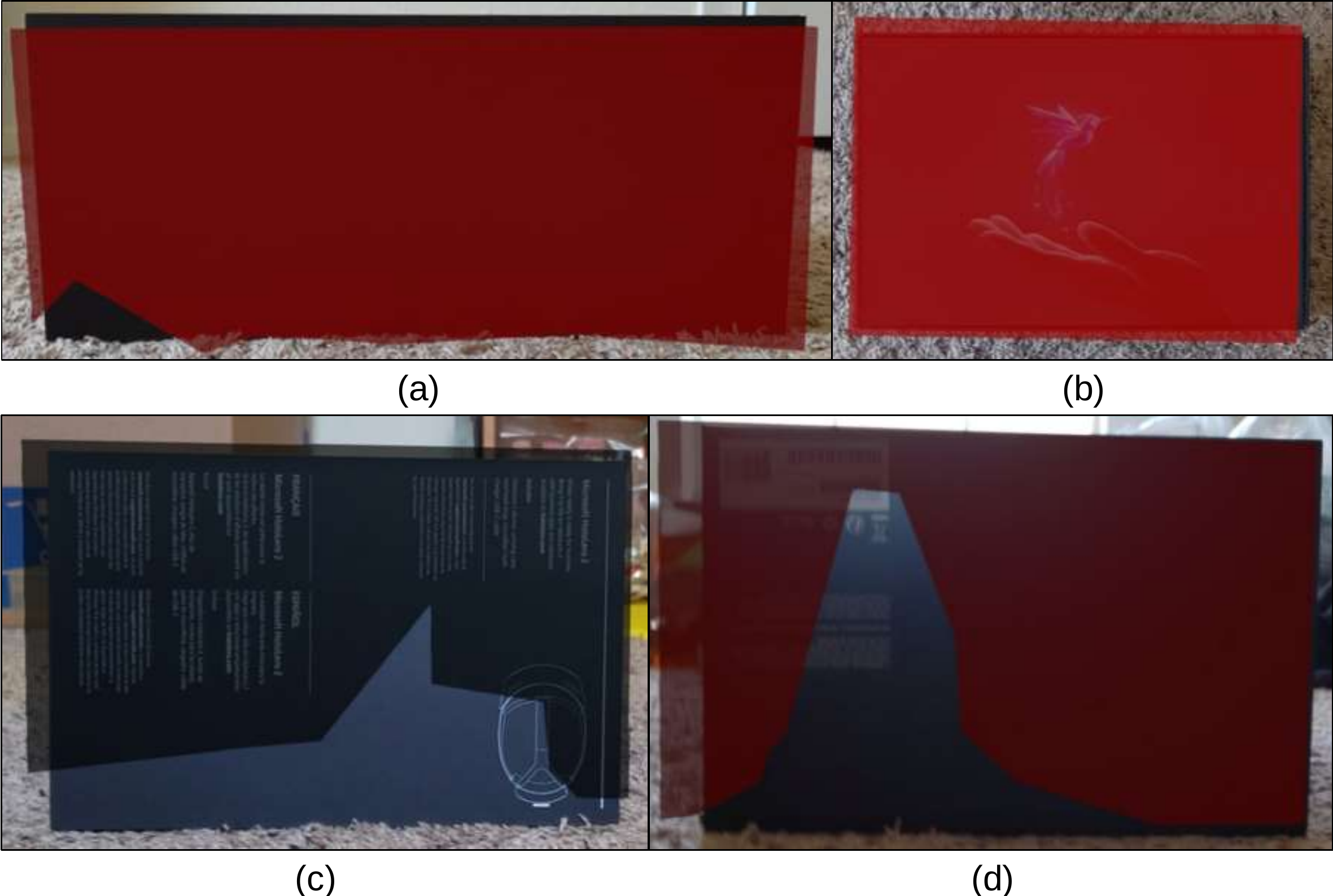}
\caption{Hologram box (red) overlap with a real-world box (black) (a) Front view (b) Top view (c) Left side view (d) Right side view}
\label{fig:hologram}
\vspace{-1em}
\end{figure}

\item Virtual and Real-world Environment Overlapping Ratio: 
This experiment indicates the flat surface overlapping of the device-generated hologram surface and the real-world target surface which it is attached. First, we create a hologram cube (0.1 m X 0.1 m X 0.1 m) in the app and ask the testers to put the cube onto the hologram surface, as shown in Figure \ref{fig:target}a. Then we will turn to the side view and take a photograph as the test result to calculate the error distance, as shown in Figure \ref{fig:target}b. We measure the gap or overlap ratio between these two surfaces by calculating the accuracy deviation \cite{Liu2018}. Figure \ref{fig:target} shows the example of the generated hologram (white line and black polygon) and the target surface (white table).

\begin{figure}[htbp]
\centering
\includegraphics[width=8.4cm, keepaspectratio=true]{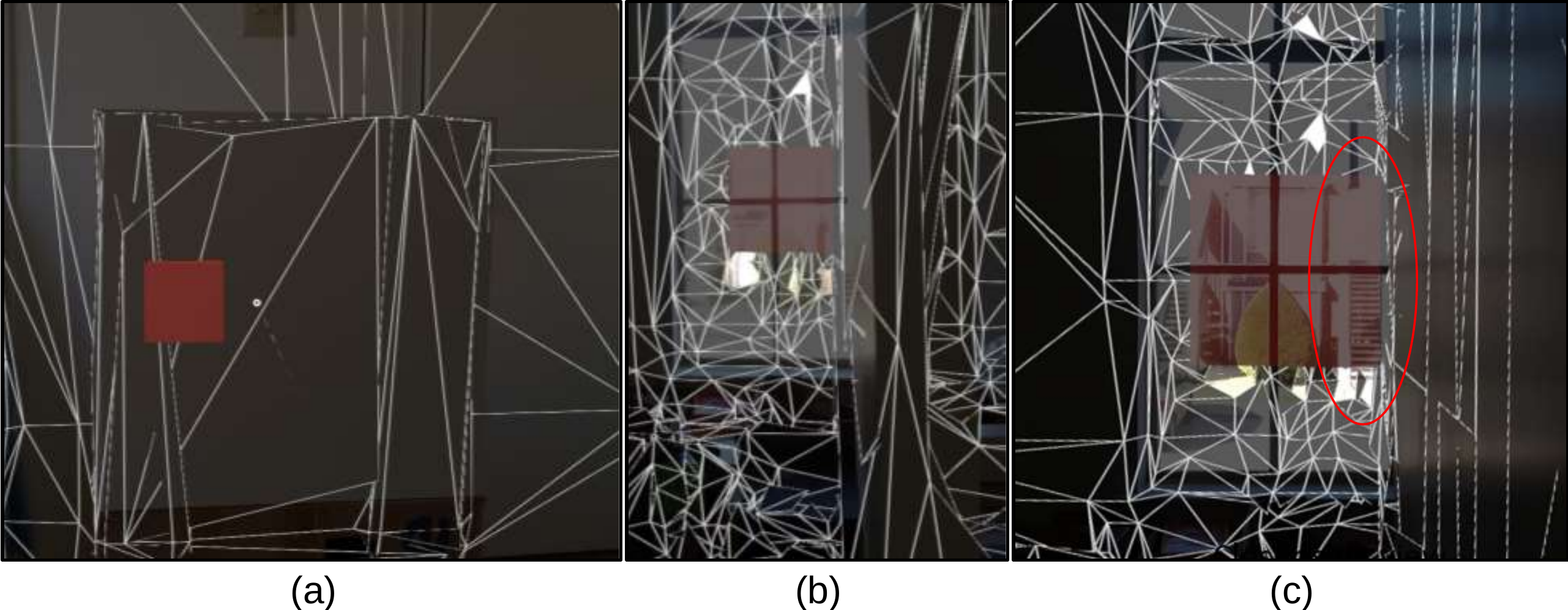}
\caption{Virtual and real-world environment overlapping process and result (a) After testers put the red virtual cube onto the hologram surface (b) Side view of the flat surface (c) Closer side view and gap example (red circle)}
\label{fig:target}
\vspace{-1em}
\end{figure}

\end{enumerate}

\subsection{Real Environment Reconstruction}
HoloLens 2 has the ability to reconstruct a real-world environment into a hologram world. In order to evaluate the reconstruction result of HoloLens 2, we use the device to generate holograms of real-world objects and indoor rooms as the evaluation output result. In this test, similar to the spatial mapping experiment, we ask the testers to remove all environmental holograms in HoloLens 2 via setting before the test start. Then the testers will act according to our instructions written as follow.

\begin{enumerate}
\item Real-world Objects Reconstruction: \hspace{\fill}\\
The real-world object reconstruction experiment is performed to evaluate the construction speed and completion percentage of HoloLens 2's object reconstruction functionality. We select four different size real-world boxes as testing objects, which are large tall box (50.2 cm X 32.5 cm X 75.5 cm), regular size box (56 cm X 47 cm X 47 cm), thin box (41 cm X 20.5 cm X 28.5 cm), small box (15.5 cm X 13.5 cm X 23 cm). \par
First, we put the four boxes in a fixed test position separately and asked the testers to open the spatial mapping app. Second, we ask the testers to look away from the box and remove all holograms via HoloLens 2's menu setting. Lastly, we instruct the testers to turn around and face the target box and walk around the box to construct the box's hologram. Then we record the construction result of the hologram for evaluation. Here, we set the construction time limit to 60 seconds; whenever the hologram becomes stable, stop recording and export the result. According to experience, a box hologram can be established stably within 60 seconds. After the box hologram becomes stable, we will ask the testers to walk around the box once again to record the final reconstruction result.
\item Real-world Indoor Room Reconstruction: \hspace{\fill}\\
In the real-world indoor room reconstruction experiment, we evaluate the performance by calculating indoor room hologram construction time and completion percentages of HoloLens 2's room reconstruction feature. In this case study, we choose four rooms, living room (7 m X 3.92 m X 2.97m), personal room (3.81 m X 3.02 m X 2.40 m), restroom (2.50 m X 2.48 m X 2.40 m), and walk-in closet (0.79 m X 0.54 m X 2.40 m) as the reconstruction target. \par
We first ask the testers to stay at a selected room corner and face the wall, then we will start recording the test process. Second, we instruct the testers to remove all holograms in HoloLens 2 and start looking around the room to construct the current nearby environment. Lastly, we ask the testers to walk along the wall with speed, 1 step (0.5 m) per second, and look around the room every step at the same time. After walking back to the original position, stop recording, and save it as the evaluation result.
\end{enumerate}

\subsection{Speech Recognition}
The speech recognition experiment is performed to evaluate the effectiveness of speech commands in HoloLens 2. To compare the speech recognition improvement from HoloLens 1 to HoloLens 2, we refer to the paper \cite{Liu2018} to select some of the testing speech commands and add some other commands. We test both system-defined commands (“select,” “Move this…there,” “face me,” “bigger/smaller,” “What's my IP address,” “hide and show hand ray,” “shut down device,” “close,” “Open the Start menu,” “Follow me,” and “take a picture”) and user-defined commands selected from Wobbrock's paper \cite{wobbrock2009user-defined} (“move,” “rotate,” “delete,” “zoom in/out,” “open,” “duplicate,” “previous,” and “help”) which mentioned in the paper \cite{Liu2018}. \par
\begin{figure}[htbp]
\centering
\includegraphics[width=8.4cm, keepaspectratio=true]{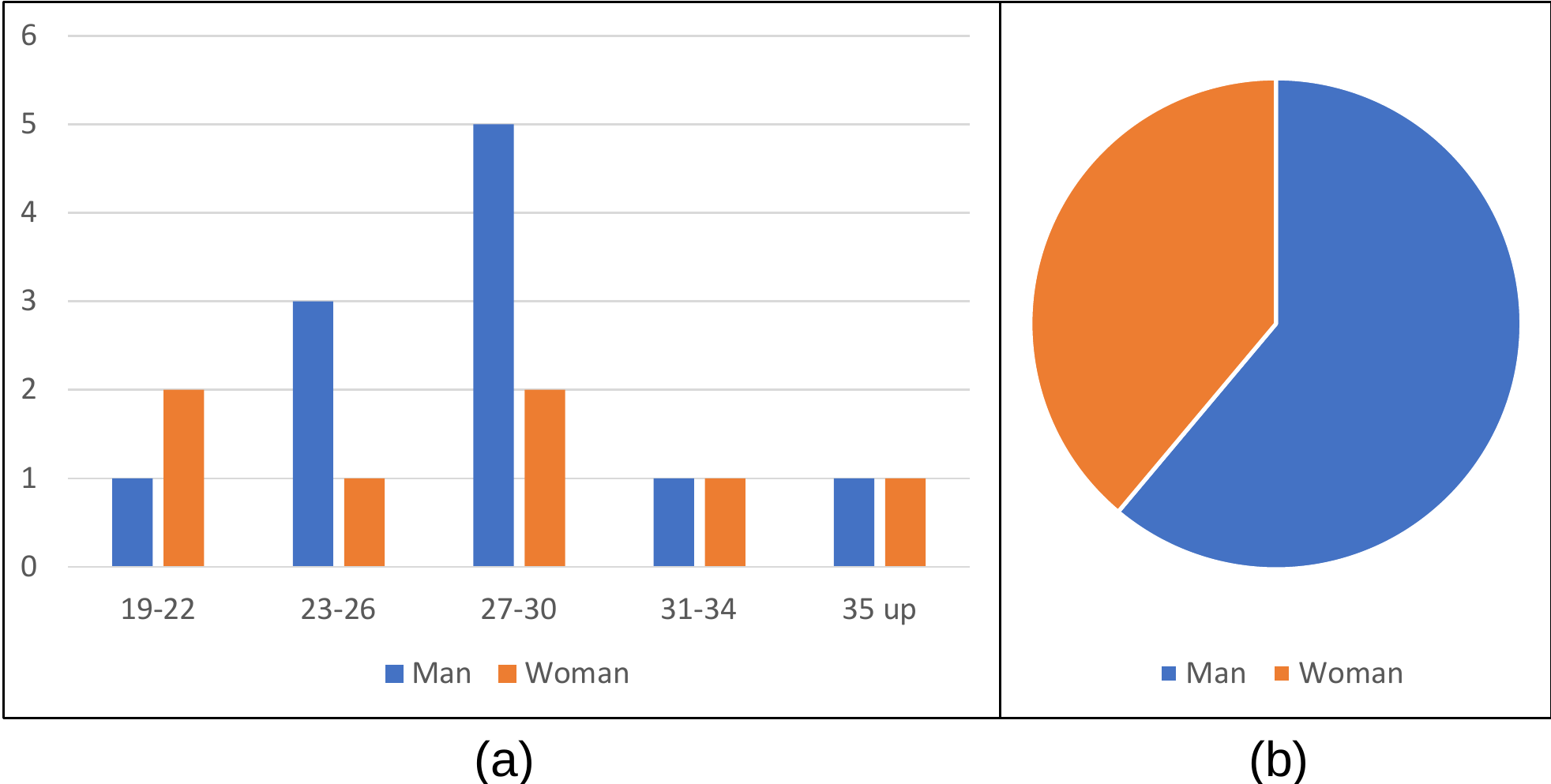}
\caption{Speech recognition native language ratio and current education status}
\label{fig:speechNative}
\vspace{-1em}
\end{figure}
\begin{figure}[htbp]
\centering
\includegraphics[width=8.4cm, keepaspectratio=true]{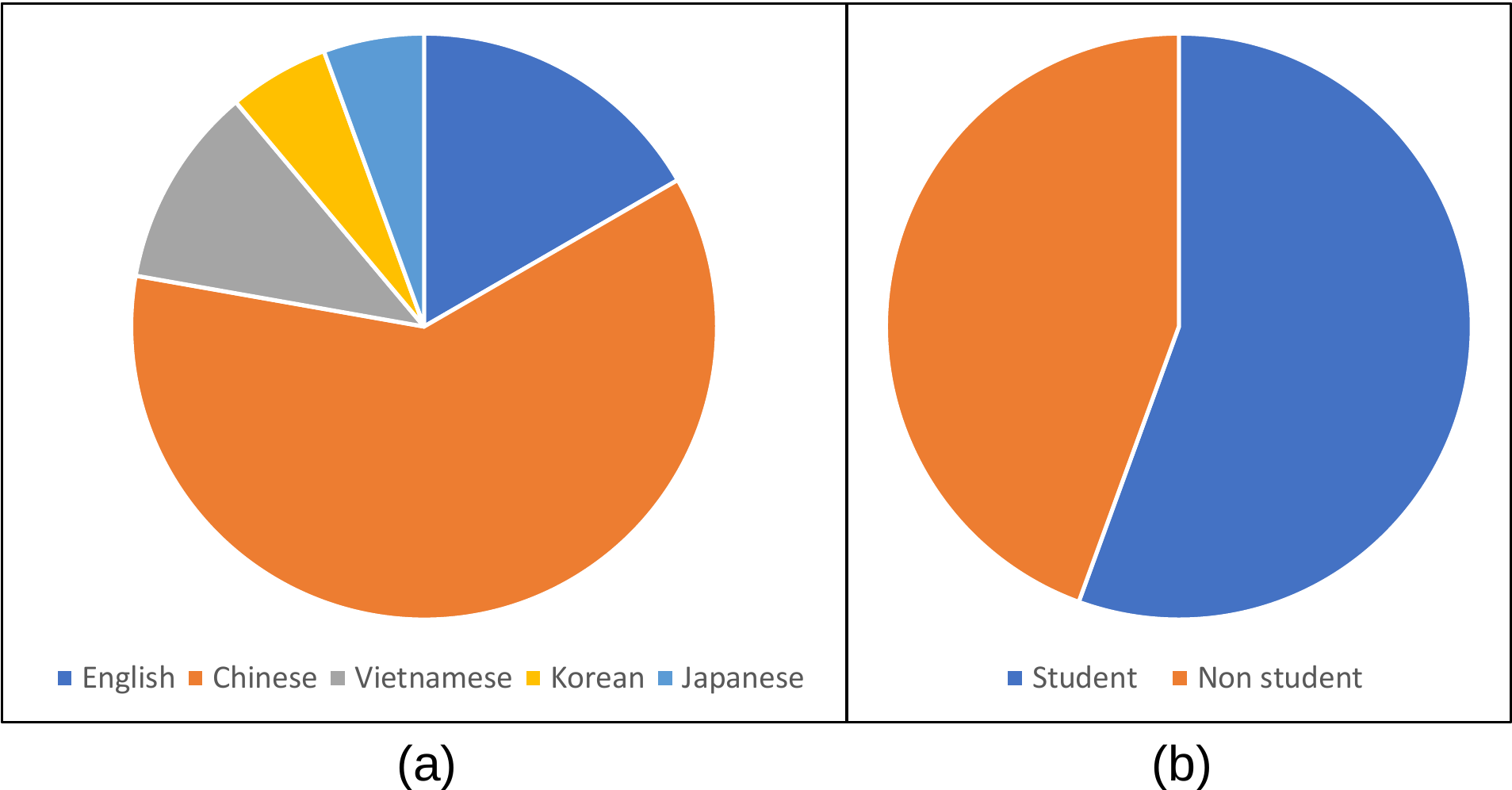}
\caption{Speech recognition gender in each age interval and total gender ratio}
\label{fig:speechGender}
\vspace{-1em}
\end{figure}

During the testing process, we ask the testers to speak each system-defined command once at a time, then open a preset app and ask the testers to speak each user-defined command once at a time. The number of recognized commands will be counted as the evaluation result to evaluate the capability of HoloLens 2's speech recognition feature. All testers lived in an English-speaking environment for more than one year and could speak English fluently. The native language proportion is shown in Figure \ref{fig:speechNative}a. We also recruited some non-student testers in order to balance the education status ratio, which is a challenging task in other MR and AR papers. The proportion of testers' current education status is shown in Figure \ref{fig:speechNative}b. Furthermore, to compare different voice line, we recruited people of varying age groups as testers with a balance in the ratio of men and women to make the experiment more accurate, as shown in Figure \ref{fig:speechGender}. 

\subsection{Azure Spatial Anchor}
Microsoft Azure provides a spatial anchors feature to create persistent, accurate digital anchors and could be operated at real-world scale. This feature can be used between HoloLen 2, Android and Apple devices. \par
HoloLens 2 uses the Azure Spatial Anchor feature to create permanent anchors in the real-world to help use the app or share world anchors with others. Currently, the best way is to use the spatial anchor feature published by Azure. For example, developers can use this feature to create a permanent world anchor arrow to guide directions for the users. \par
\begin{figure}[htbp]
\centering
\includegraphics[width=8.4cm, keepaspectratio=true]{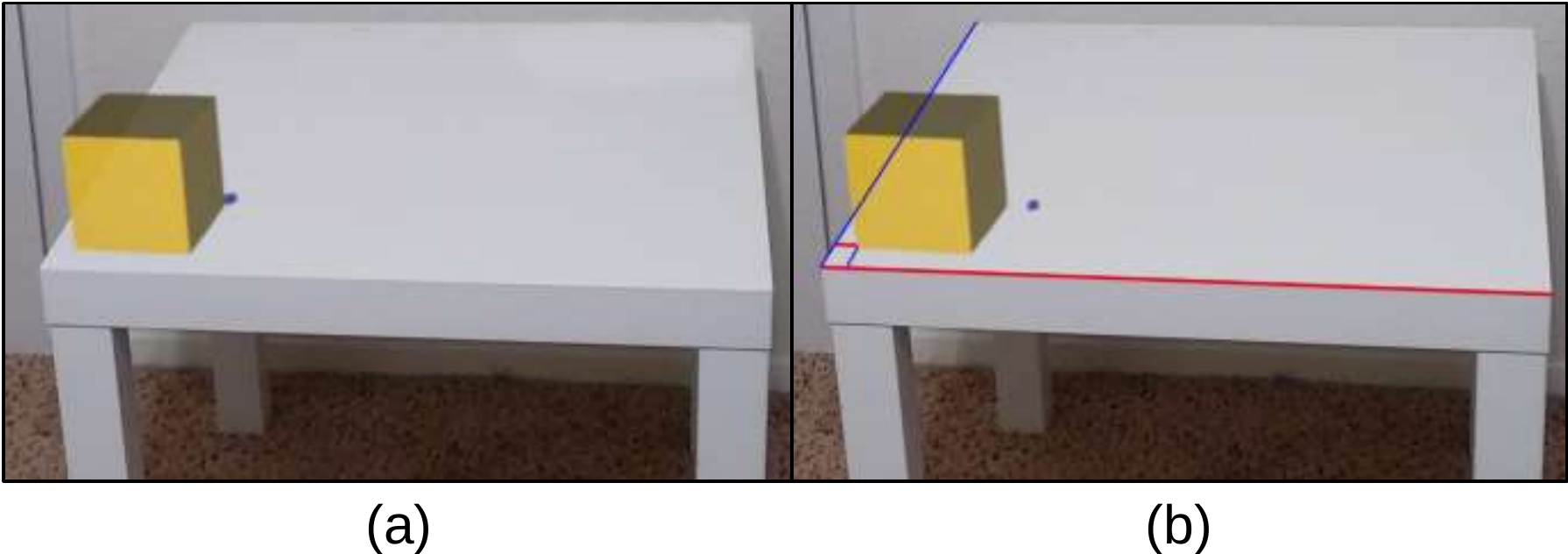}
\caption{Azure Spatial Anchor on a table in the real-world (a) User-created benchmark on the local end (b) User-created anchor download from the server}
\label{fig:Azure}
\vspace{-1em}
\end{figure}

Therefore, we carry out HoloLens 2 Azure Spatial Anchor position evaluation to determine whether the position, in real-world, has changed from uploading the spatial anchors to the cloud to downloading back to the local device. First, we need to connect the HoloLens 2 app with the Azure Spatial Anchor server. Then we ask the testers to start the Azure session and create an anchor cube on the real-world table and take a picture as a test benchmark. Here, the anchor's position is saved to the Azure Spatial Anchor server, as shown in Figure \ref{fig:Azure} (a). Next, the testers move the cube to another place and locate the anchor created previously from the server by pressing a button in the app; the downloaded cube shows in Figure \ref{fig:Azure}b. Finally, take a picture for the Anchor and calculate the error distance between the benchmark and the downloaded result.

\subsection{Research Mode -- IMU sensors}
In the HoloLens 2 research mode \cite{ungureanu2020hololens}, one of the most significant difference from the previous HoloLens 1 is the addition of accessing the inputs of Inertial Measurement Unit (IMU) sensors, Accelerometer, Gyroscope, and Magnetometer. This can help researchers or users gain more understanding of machine movement or rotation. This subsection focuses on using Accelerometer and Gyroscope to calculate the error between reality and the device's output. Then will describe the detailed design of testing two components of the IMU sensors in HoloLens 2 \cite{ungureanu2020hololens}. Since there is a lot of external magnetic field interference in the test environment, we are not doing Magnetometer-related experiments here. 
\begin{enumerate}
\item Accelerometer: \hspace{\fill}\\
The Accelerometer in HoloLens 2 \cite{ungureanu2020hololens} is used to determine linear acceleration along the X, Y, and Z axes and gravity. The output data from Accelerometer is the projection of the force on X, Y, and Z axes of the acceleration, combined acceleration, and time. Therefore, by using this component, we can compute the users' moving direction and distance. Also, the projection of the acceleration force can be used to determine the tilt angle of the users. In this experiment, we use these two measurements to ask the testers to move a fixed distance along the x, y, or z-axis or tilt their body in a specified direction to test the accuracy. \par
Since the Accelerometer provides the linear acceleration along the X, Y, and Z axes, we can use double integrals to obtain the current position. According to the output data from the whole IMU sensors, we know that the HoloLens 2 uses a three-dimensional space setting, as shown in Figure \ref{fig:IMU}b. On the basis of the outputs from the Accelerometer, let \(r_{x}\), \(r_{y}\), \(r_{z}\) representing the projection of the force on X, Y, and Z axes of the acceleration, which show the instantaneous acceleration on the current point. Since the time interval between each acceleration point is 0.083 seconds, which is very short, we can assume that the instantaneous acceleration is equal to the average acceleration between each point. Let one of the average acceleration \(a_{1} = \left ( a_{1}x, a_{1}y, a_{1}z \right )\), and another average acceleration \(a_{2} = \left ( a_{2}x, a_{2}y, a_{2}z \right )\) right after \(a_{1}\). Then, we may assume a linear equation \(\mathbf{a}(t)\) to represent points between \(a_{1}\) and \(a_{2}\) as
\begin{equation}
\begin{aligned}
\mathbf{a}(t) = (a_{1}x + (a_{2}x - a_{1}x) * 12t, \\
a_{1}y + (a_{2}y - a_{1}y) * 12t, \\
a_{1}z + (a_{2}z - a_{1}z) * 12t).
\end{aligned}
\end{equation}
Since we assume that the instantaneous acceleration is the same as average acceleration, we can integrate \(\mathbf{a}(t)\) to yield velocity \(\mathbf{v}(t)\)
\begin{equation}
\begin{aligned}
\mathbf{v}(t) = (a_{1}xt + (a_{2}x - a_{1}x) * 6t^{2} + C_{x}, \\
a_{1}yt + (a_{2}y - a_{1}y) * 6t^{2} + C_{y}, \\
a_{1}zt + (a_{2}z - a_{1}z) * 6t^{2} + C_{z}).
\end{aligned}
\end{equation}
Here \(C_{x}\), \(C_{y}\), and \(C_{z}\) are the constants which represent the velocity calculated previously at the point \(a_{1}\) and we assume first point’s velocity is \((0,0,0)\). Secondly, we can integrate \(\mathbf{v}(t)\) to produce position \(\mathbf{s}(t)\)
\begin{equation}
\begin{aligned}
\mathbf{s}(t) = (\frac{1}{2}a_{1}xt^{2} + (a_{2}x - a_{1}x) * 2t^{3} + C_{x}t + K_{x}, \\
\frac{1}{2}a_{1}yt^{2} + (a_{2}y - a_{1}y) * 2t^{3} + C_{y}t + K_{y}, \\
\frac{1}{2}a_{1}zt^{2} + (a_{2}z - a_{1}z) * 2t^{3} + C_{z}t + K_{z}).
\end{aligned}
\end{equation}
Here \(K_{x}\), \(K_{y}\), and \(K_{z}\) are the constants which represent the position calculated previously at the point \(a_{1}\) and we also assume first point’s position is \((0,0,0)\). Lastly, we can use \(\mathbf{s}(t)\) calculated on every two points \(a_{1}\) and \(a_{2}\) to calculate the current position of HoloLens 2, and compute the moving distance accordingly to compare with the actual moving distance. 

\begin{figure}[htbp]
\centering
\includegraphics[width=8.4cm, keepaspectratio=true]{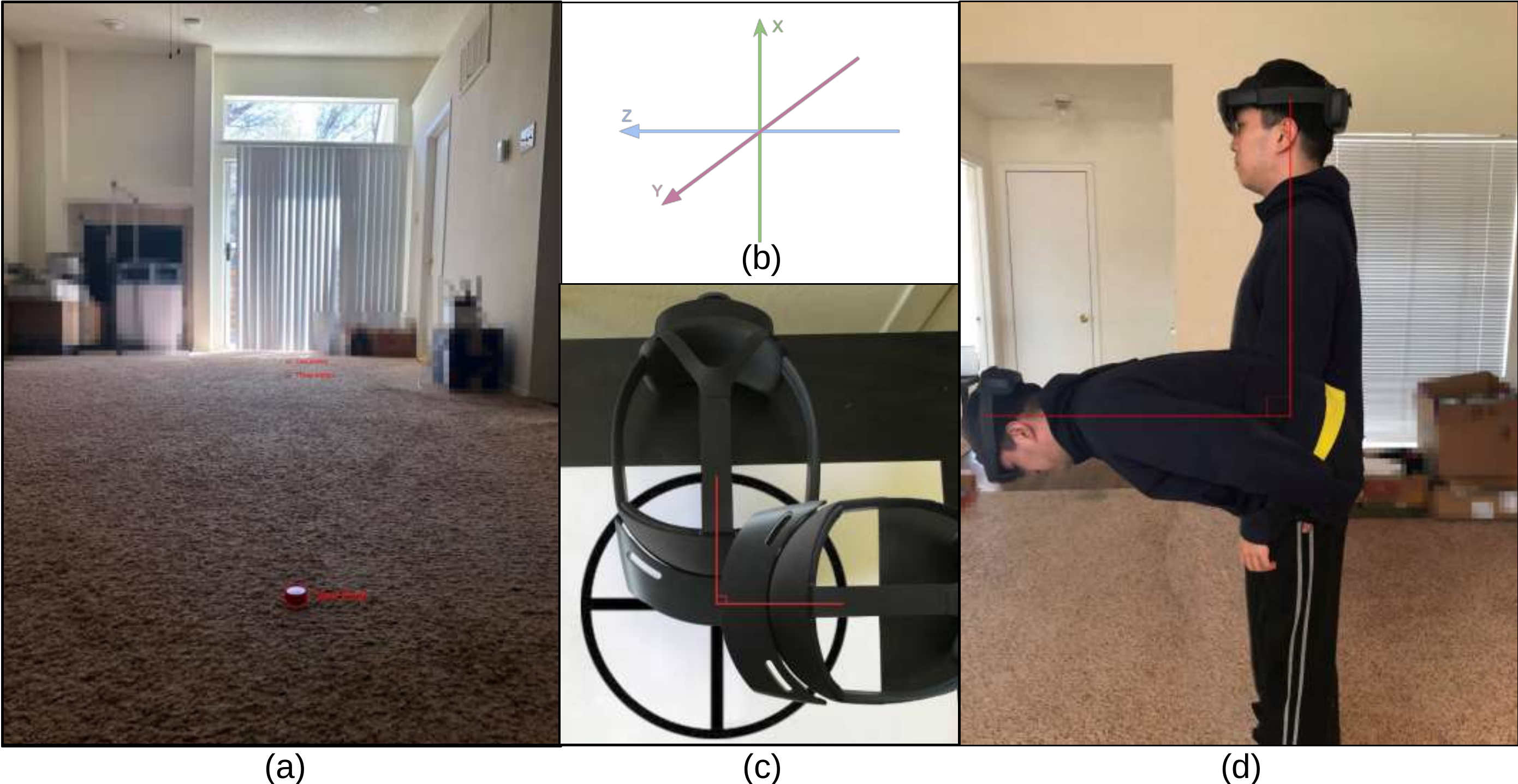}
\caption{(a) Accelerometer walking distance setting (b) Three dimensional space setting used in HoloLens 2 (c) Example of Gyroscope rotating 90 degree along x-axis (d) Example of Accelerometer tilt 90 degree along x-axis and z-axis}
\label{fig:IMU}
\vspace{-1em}
\end{figure}

In addition, we also use the output data from Accelerometer to evaluate the tilting angle. As shown in Figure \ref{fig:IMU}d, we ask the testers to tilt their head at a fixed angle when wearing the HoloLens 2 device. Then according to the output \(r_{x}\), \(r_{y}\), \(r_{z}\), we can use the following equation to calculate the tilt angle along each axes.
\begin{equation}
\begin{aligned}
A_{xr} = arccos(\frac{R_{x}}{R}) * (\frac{180}{\pi})
\end{aligned}
\end{equation}
where \(R= \sqrt{R_{x}^{2} + R_{y}^{2} + R_{z}^{2}}\) to express the combined force on the current location. Since the data output from the device is in the form of radian by default, we need to multiply \(\frac{180}{\pi}\) to transform it into degree form.

\item Gyroscope: \hspace{\fill}\\
Using the Gyroscope, we can extract the output data, including angular velocity on X, Y, and Z axes and time. According to these data, we can determine the rotation angle made by the users. In order to achieve accurate measurement, we use computer software to draw a circle and a vertical cross and print them on paper as ground truth with 90-degree intervals between each line, as shown in Figure \ref{fig:IMU}c. We can determine the computation error in this evaluation process by comparing the device's rotation angle and the calculated rotation angle. \par
According to the Gyroscope output data from the HoloLens 2, we can find out that there are three angular velocities per unit of time, \(G_{x}\), \(G_{y}\), and \(G_{z}\), and each unit of time is 0.048 seconds. Therefore, we can transform the three angular velocities to actual angle change per unit time by substituting the following formula.
\begin{equation}
\begin{aligned}
AC = ((G_{x}*\frac{180}{\pi}) * 0.048, \\
(G_{y}*\frac{180}{\pi}) * 0.048, \\
(G_{z}*\frac{180}{\pi}) * 0.048)
\end{aligned}
\end{equation}
Similarly, the data output from the device is in the form of radian by default, so we need to multiply \(\frac{180}{\pi}\) to transform it into degree form. By calculating angle changes for every unit of time, we can compare the output of the Gyroscope with the actual rotation angle.

\end{enumerate}



\section{Result and Discussion}
According to the previous experiment design and procedure sections, this section contains the experimental results and discussions about the outcome of the results. 

\subsection{Hand Tracking}
For the hand tracking experiment, we calculated the error distance between each hologram hand joint point (Figure \ref{fig:joint}a) and real-world hand joint point (Figure \ref{fig:joint}b) in the five frames cropped from the recorded experiment video. The error distance \textit{ER} is calculated as follow:
\begin{equation}
\begin{aligned}
ER = \frac{1}{N}\sum_{i\in N}^{}d\left ( r_{i},g_{i} \right )\qquad
\end{aligned}
\end{equation}
where \textit{N} represents the number of hand joints, \textit{d} is the Euclidean distance of two points, and \(r_{i}\), \(g_{i}\) are the corresponding hand joint point in real-world hand and HoloLens 2 generated hand. \textit{ER} is greater than or equal to 0, where 0 means that the points generated by HoloLens 2 are closer to the real points, and higher {ER} represents the larger the error is.
\par
\begin{figure}[htbp]
\centering
\includegraphics[width=8.4cm, keepaspectratio=true]{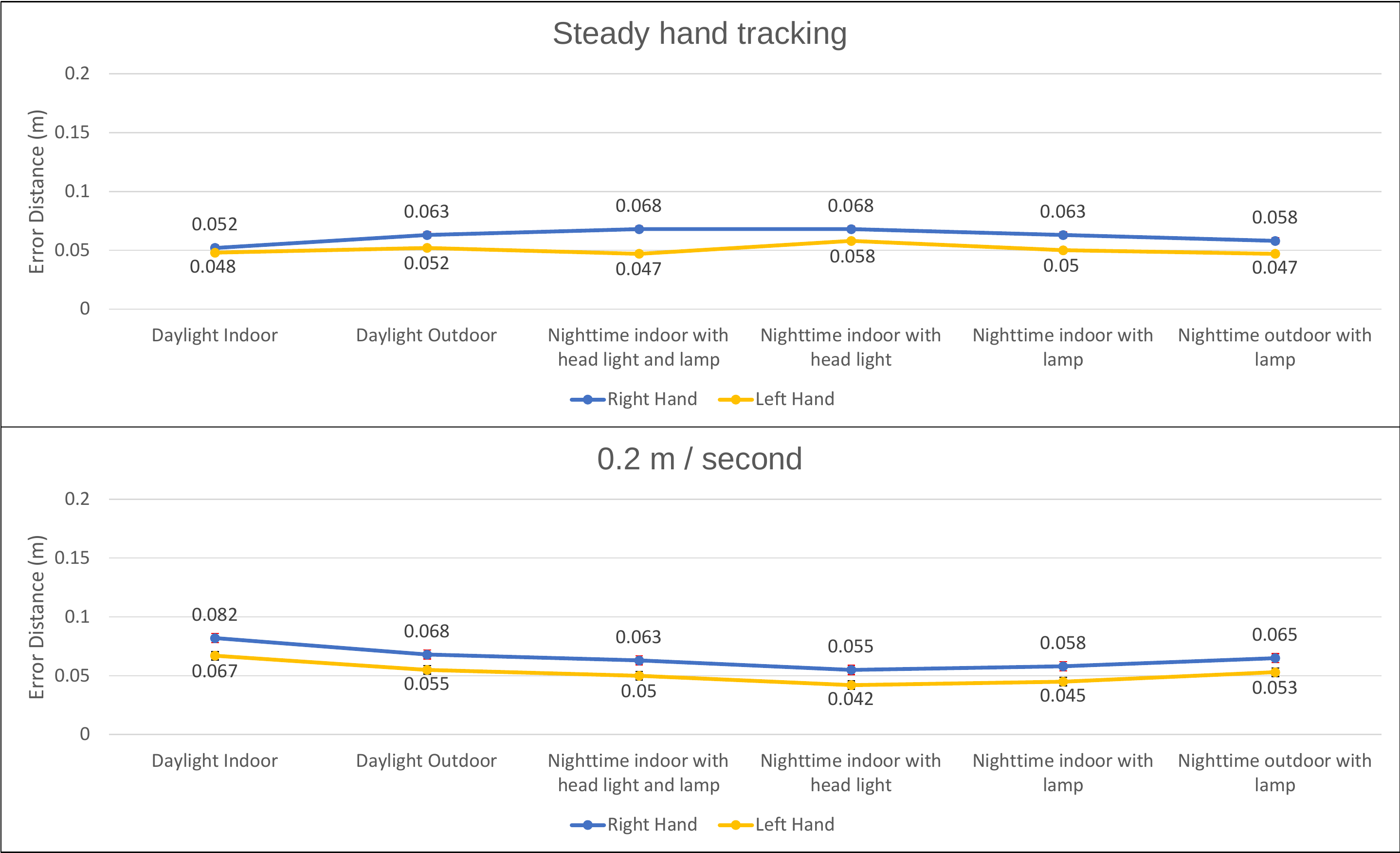}
\includegraphics[width=8.4cm, keepaspectratio=true]{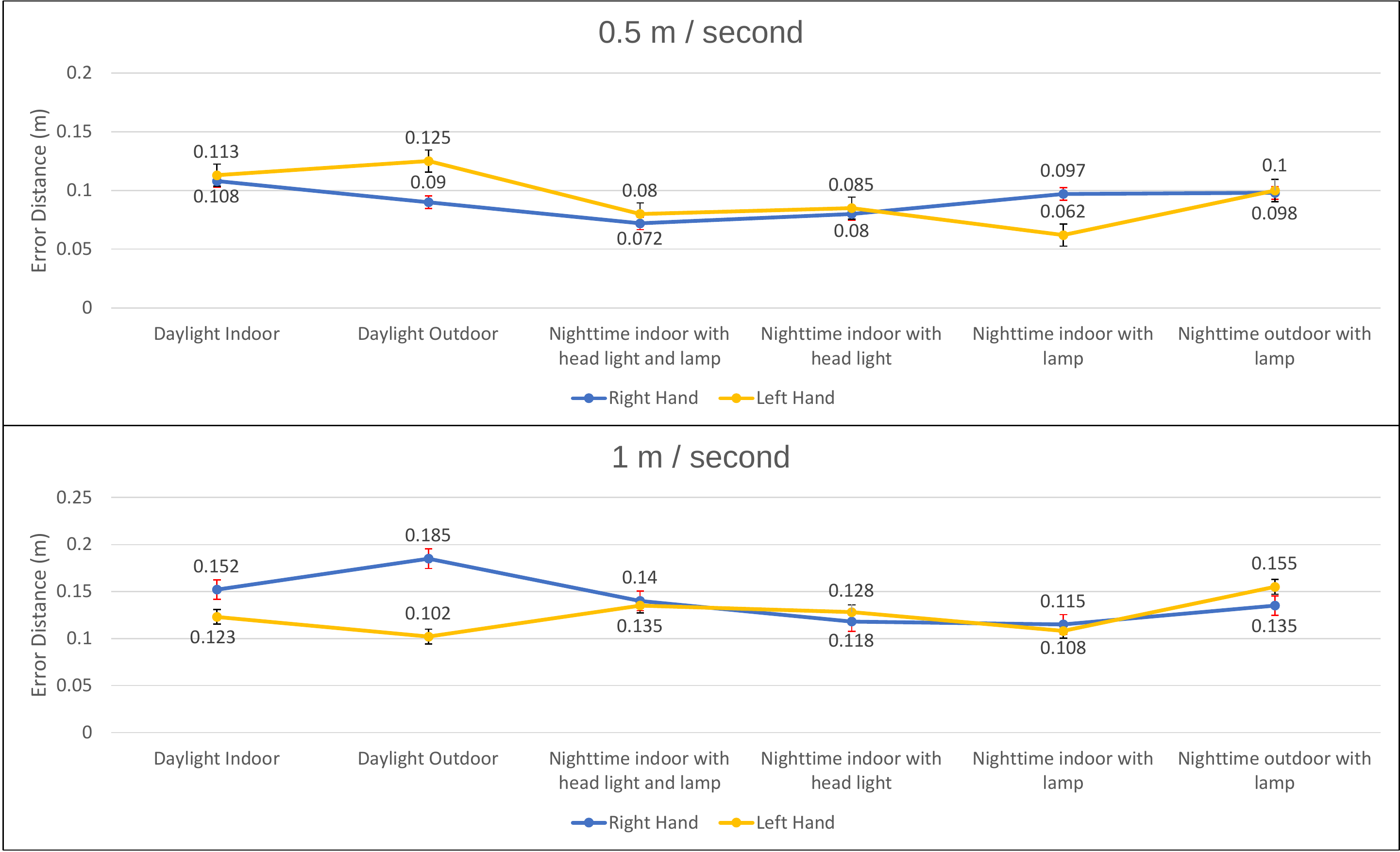}
\caption{Hand tracking experiment result -- error distance with standard error bars (red: right hand, black: left hand) under each speed and environmental conditions}
\label{fig:handResult}
\vspace{-1em}
\end{figure}
\begin{figure}[htbp]
\centering
\includegraphics[width=8.4cm, keepaspectratio=true]{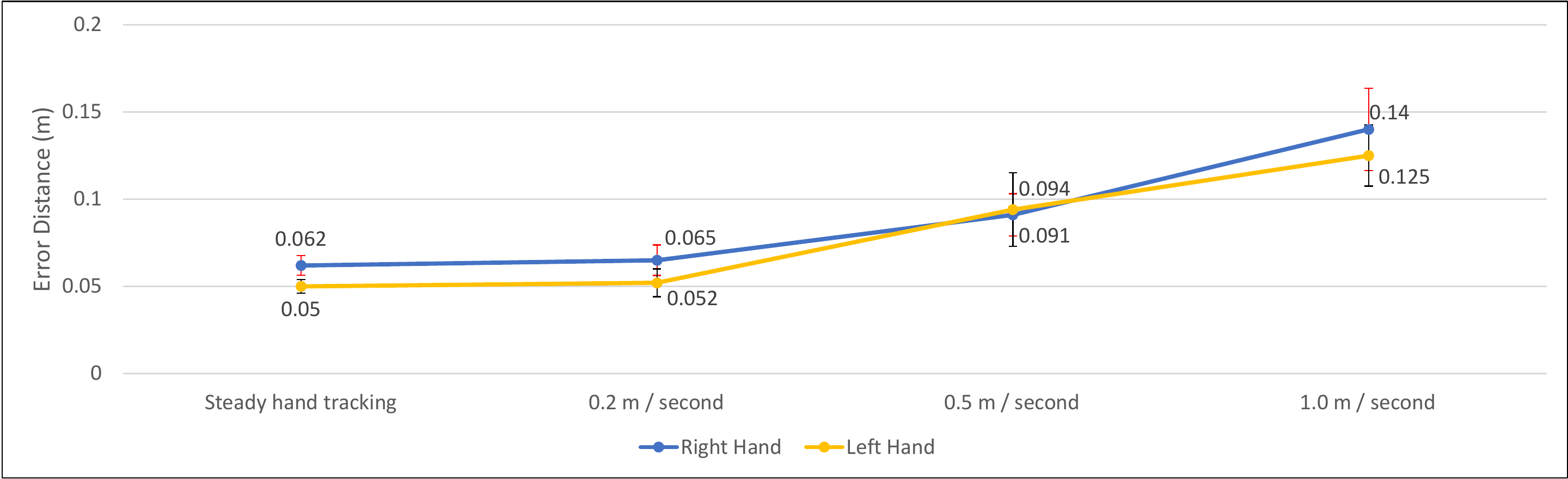}
\caption{Hand tracking experiment average error distance with standard error bars under each speed conditions}
\label{fig:handError}
\vspace{-1em}
\end{figure}

We split the experiment result into four different speed conditions, steady, 0.2 m / sec, 0.5 m / sec, and 1 m / sec, as shown in Figure \ref{fig:handResult}. Also, we calculate the average error distance of all environmental conditions of each speed condition and organize it in a chart, which shows in Figure \ref{fig:handError}. 

\par
Although the error distance seems a little bit too high in the mixed reality world in steady hand tracking, the hands' hologram in the user's eyes still seems precisely match the user's real hands. Therefore, the actual error distance under each speed conditions can be computed relative to the steady hand tracking error distance result. According to the average error distance result, we can find that the error distance will be more considerable when the speed is faster. However, the hand tracking error distance is not significant even at 1 m / sec speed; we can say that this result is quite ideal below this speed. Thence, when developers or users try to develop a new application in HoloLens 2, we recommend them to be aware of hands moving speed not to go over 1.0 m / sec when moving from one scene to the other.

\subsection{Eye Tracking}
In the eye-tracking experiment result, after collecting the eye-tracking experiment result images, we calculated the average error distance under different environmental conditions. 
\begin{enumerate}
\item Virtual Objects Eye Gaze Tracking: \hspace{\fill}\\
In this experiment result, we calculate the error distance using the same error distance \textit{ER} in the previous section between five virtual cubes and corresponding eye gaze points. Then, we compute the average of the error distances under each position and environmental conditions, as shown in Figure \ref{fig:eyeError}(top).\par
The eye-tracking error distance results show every error distance is smaller than 5 centimeters. In the Night-time indoor with lamp only experiment, the lamplight is from the user's back to the center-top side; therefore, the error distance in the top and center is smaller than the other side. Similarly, in the Night-time outdoor with lamp only experiment, the lamplight is from the user's left side; thus, the left side's error distance is smaller than the other left side results at night.
\item Real-world Objects Eye Gaze Tracking: \hspace{\fill}\\
In the real-world objects eye gaze tracking experiments, we calculate the average error distance \textit{ER} between five virtual cubes and corresponding eye gaze points under two environmental conditions, as shown in Figure \ref{fig:eyeError}(bottom).\par
The bar graph shows the average error distance under two environmental conditions; the results are similar to the above eye gaze tracking results, which means that eye-tracking results can achieve good results both in virtual and real environments. 

\begin{figure}[htbp]
\centering
\includegraphics[width=8.4cm, keepaspectratio=true]{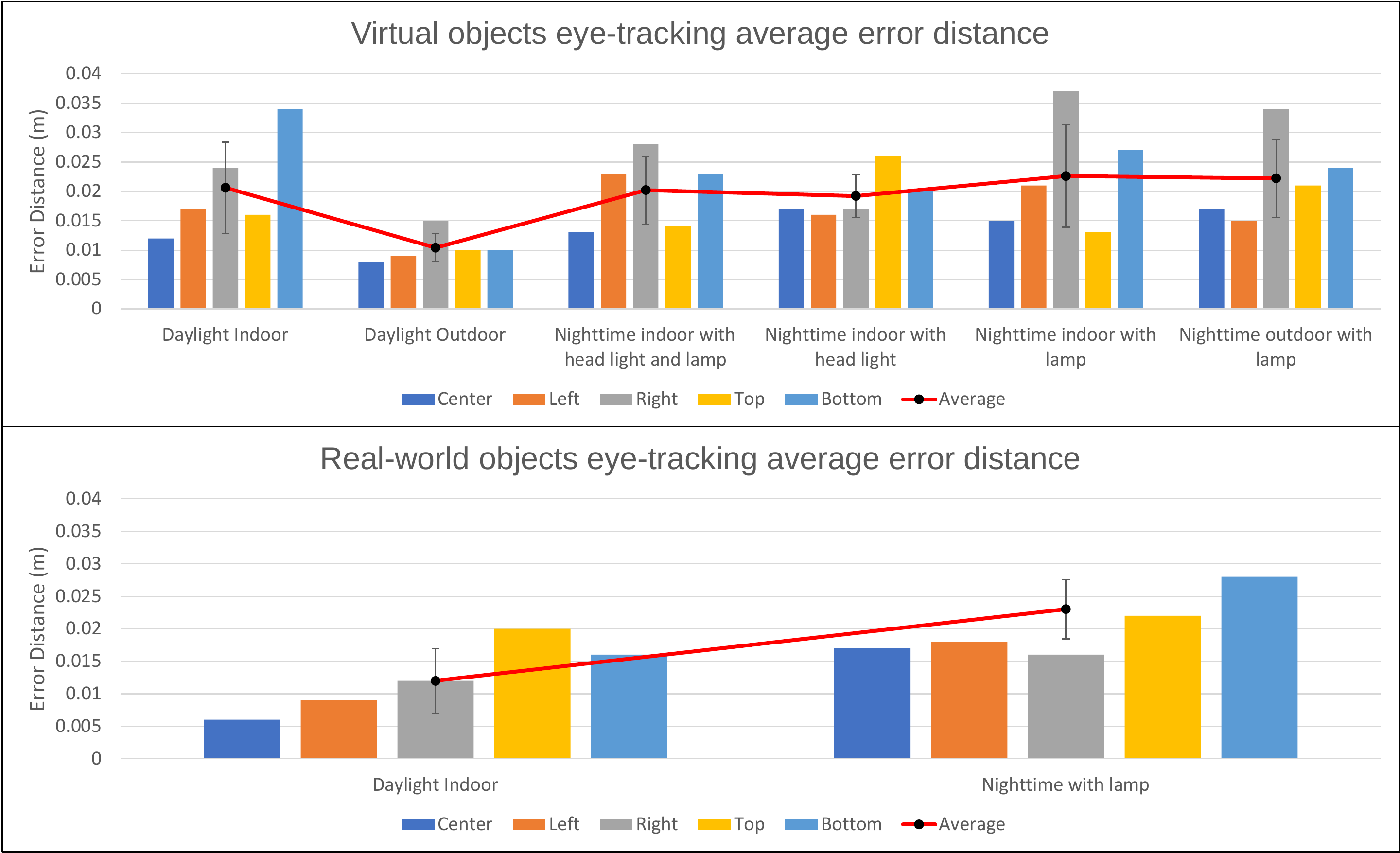}
\caption{Eye-tracking average error distance with standard error bars on virtual objects (top) and real-world objects (bottom)}
\label{fig:eyeError}
\vspace{-1em}
\end{figure}

\end{enumerate}

\subsection{Spatial Mapping}
In spatial mapping experiments, we compare with the experimental results from HoloLens 1 \cite{Liu2018}. For this comparison, we choose to use the accuracy deviation \(\sigma_{A}^{R}\), as in \cite{Liu2018}. The accuracy deviation \(\sigma_{A}^{R}\) is used to compare the difference between real-world environment and its reconstructed model: 
\begin{equation}
\begin{aligned}
\sigma_{A}^{R} = \sum_{i\in  N}^{} \left ( \frac{\left | L - l_{i} \right |}{L\cdot N} \right )\qquad
\end{aligned}
\end{equation}
where \textit{N} represents the number of measurements, \textit{L} refers to the real-world object's length of one edge, and \(l_{i}\) is the corresponding edge length measured in the \textit{i}th measurement by HoloLens 2. The value of \(\sigma_{A}^{R}\) is greater than or equal to 0, where 0 represents that the real-world objects and their holographic model generated by HoloLens 2 are identical, and higher \(\sigma_{A}^{R}\) value means there is significant difference between the two.

\begin{enumerate}
\item Hologram Visualization: \hspace{\fill}\\
For the hologram visualization experiments, we first calculate the length, width, and height of the hologram box created by our app and the real-world box. Then, we calculate the difference between each edge of real-world box and hologram box using accuracy deviation \(\sigma_{A}^{R}\). The result shows in Figure \ref{fig:holoVisual}. Since each view has an edge that cannot be calculated, we only calculate two edges' accuracy deviation on each view. This situation can be seen in Figure \ref{fig:hologram}. 

\begin{figure}[htbp]
\centering
\includegraphics[width=8.4cm, keepaspectratio=true]{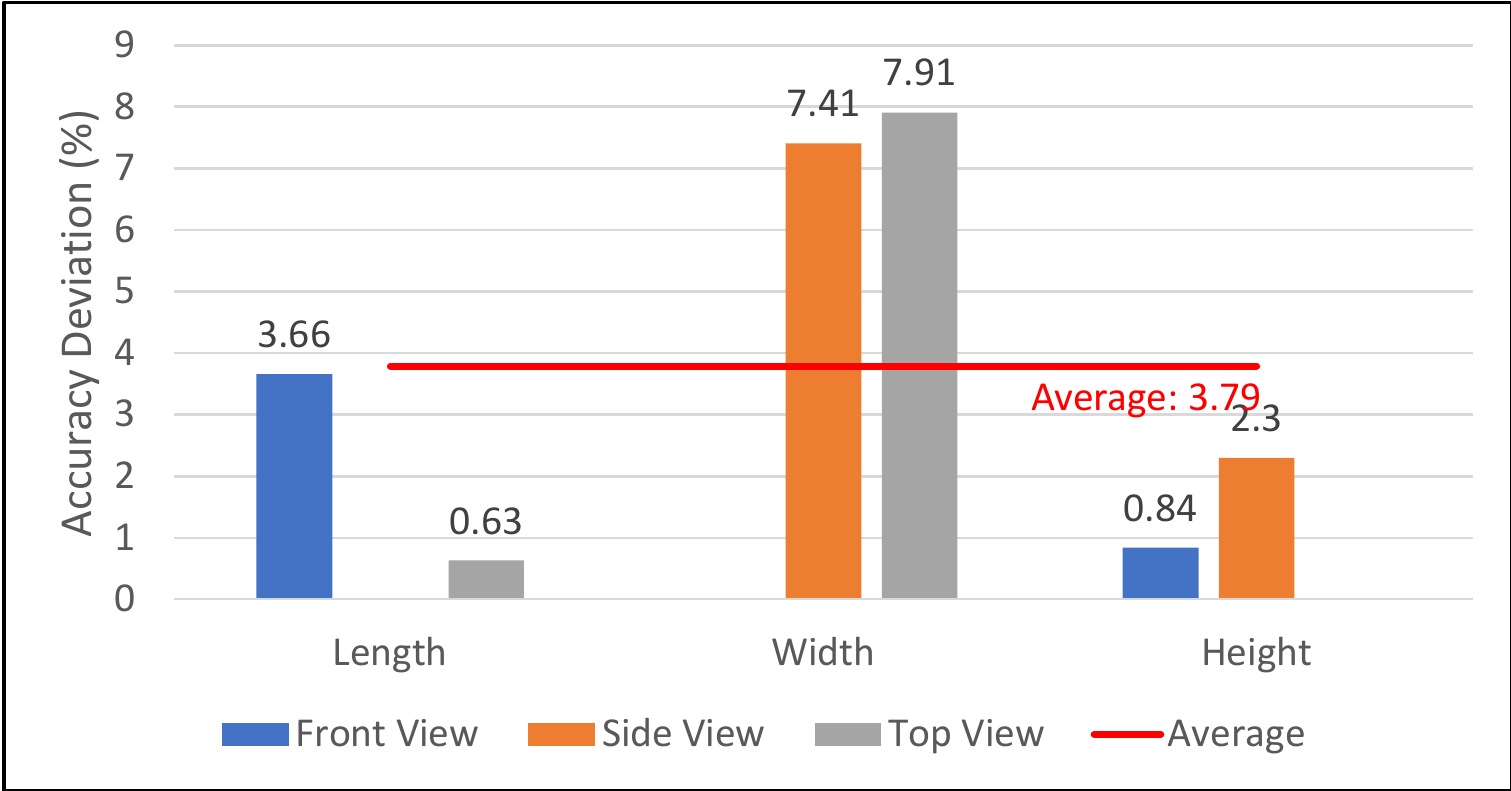}
\caption{Hologram visualization accuracy deviation on each edge and view}
\label{fig:holoVisual}
\vspace{-1em}
\end{figure}

As we can see from Figure \ref{fig:holoVisual}, HoloLens 2 performs really well on each view and edge, especially the length and height on each view. In the previous paper, the average accuracy deviation of HoloLens 1's hologram visualization result is 6.64\%, which is higher than the HoloLens 2's average accuracy deviation 3.79\% in this paper. 

\item Virtual and Real-world Environment Overlapping Ratio: 
For this experiment, we first calculate the gap or overlap between the red cube and the real-world flat surfaces. To compare with the previous HoloLens 1 paper, we use accuracy deviation \(\sigma_{A}^{R}\) to evaluate the result. Where \textit{L} refers to the real-world length of hologram cube's one edge, and \(l_{i}\) is the distance between the distance from the farthest edge to the target surface. However, we do not use the same experiment. Instead, we use different environmental conditions as our environmental variables. The result shows in Figure \ref{fig:overlappingRatio}.

\begin{figure}[htbp]
\centering
\includegraphics[width=8.4cm, keepaspectratio=true]{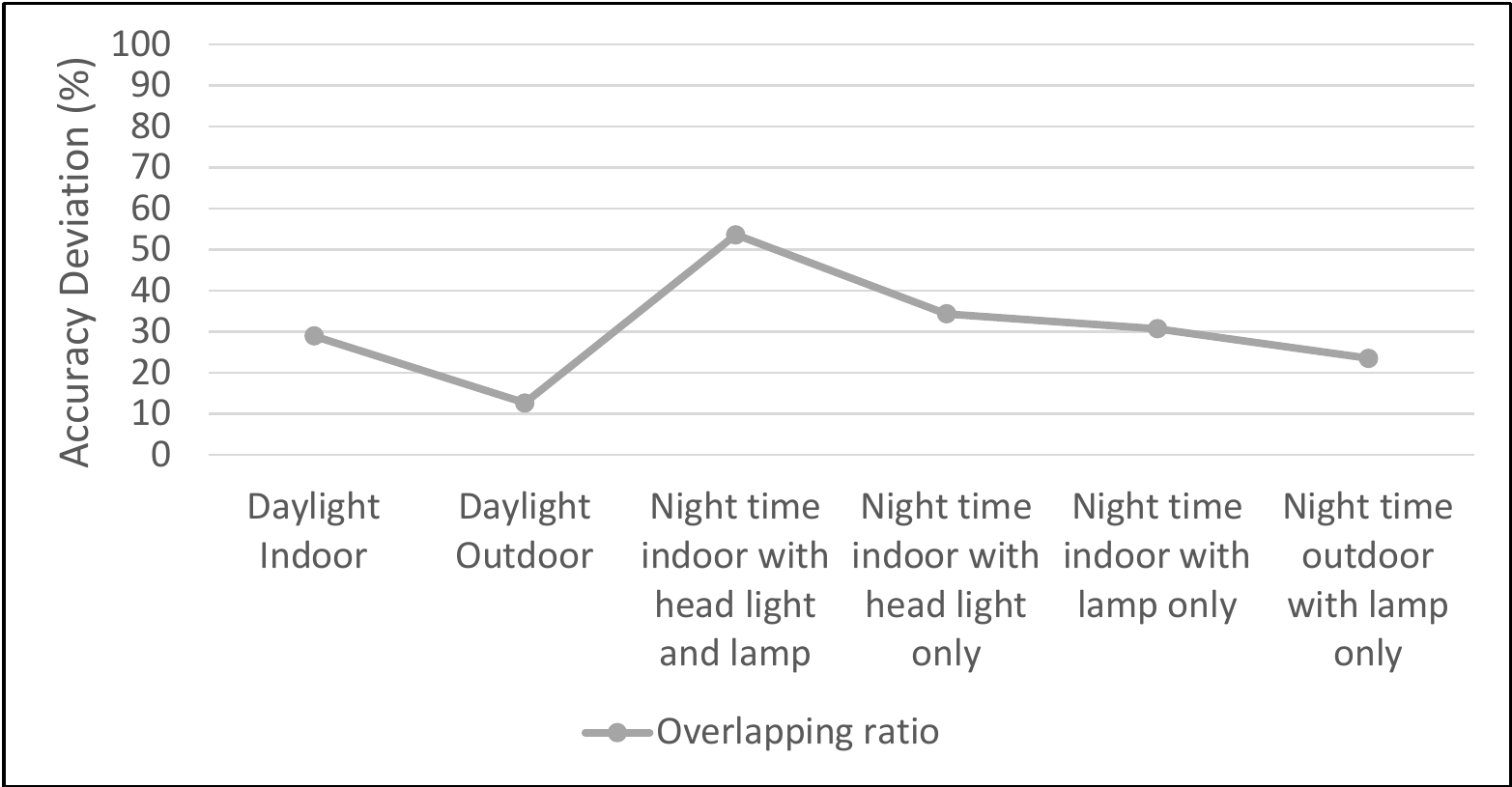}
\caption{Virtual and real-world environment overlapping ratio result}
\label{fig:overlappingRatio}
\vspace{-1em}
\end{figure}

In the HoloLens 1 environment overlapping ratio, the best accuracy deviation is 70\%, however, in the HoloLens 2 environment overlapping ratio result, the lowest accuracy deviation is 53.69\%, and the average accuracy deviation is 30.71\%, which is much better than the previous device. This result shows excellent progress from HoloLens 1 to HoloLens 2 on reconstructing the real-world environment. \par
Furthermore, according to our experiment result, we recommend users or developers to have enough illumination on the target surface to generate better hologram results.

\end{enumerate}

\subsection{Real Environment Reconstruction}
In the real environment reconstruction experiment result, after recording the reconstruction process video, we estimate the construction result using two metrics: construction time and completion percentage. The completion percentage \textit{CP} is calculated as follows:
\begin{equation}
\begin{aligned}
CP = \frac{\sum_{i\in N}^{} H_{i}}{\sum_{i\in N}^{} R_{i}}\qquad
\end{aligned}
\end{equation}
where \textit{N} represents the number of measurements, \(H_{i}\) is the hologram area generated by HoloLens 2, which is estimated by using the square method, and \(R_{i}\) is the real environment area.

\begin{enumerate}
\item Real-world Objects Reconstruction: \hspace{\fill}\\
In this experiment, we first record the construction time, then we use the reconstruction result frames to calculate the construction completion percentage, \textit{CP}. The \(H_{i}\) here means the hologram area on each side of the box, and \(R_{i}\) is the area of the real-world box. The result shows in Figure \ref{fig:realReconstruct}.

\begin{figure}[htbp]
\centering
\includegraphics[width=8.4cm, keepaspectratio=true]{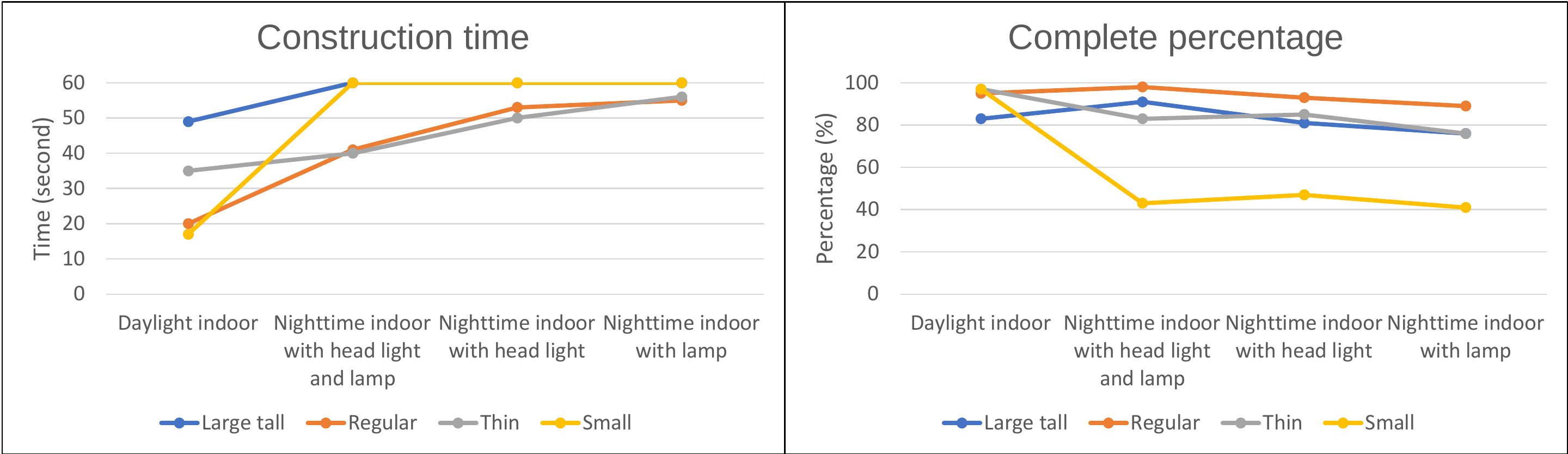}
\caption{Real-world objects reconstruction time and complete percentage}
\label{fig:realReconstruct}
\vspace{-1em}
\end{figure}

This experiment is used to explore how much object size will affect objects construction outcomes in HoloLens 2. According to the result, object size and illumination will affect the construction result and the construction time. The larger the real-world box, the longer time will take for HoloLens 2 to construct the result. However, if the real-world box is smaller than the small box and has insufficient illumination, the HoloLens 2 will see this object as an environment obstruction or ignore it and fail to construct a good hologram.

\item Real-world Indoor Room Reconstruction: \hspace{\fill}\\
Similar to the real-world objects reconstruction experiment, in this experiment, we document the construction time and calculate the construction complete percentage according to the recorded video, as shown in Figure \ref{fig:realComplete}.

\begin{figure}[htbp]
\centering
\includegraphics[width=8.4cm, keepaspectratio=true]{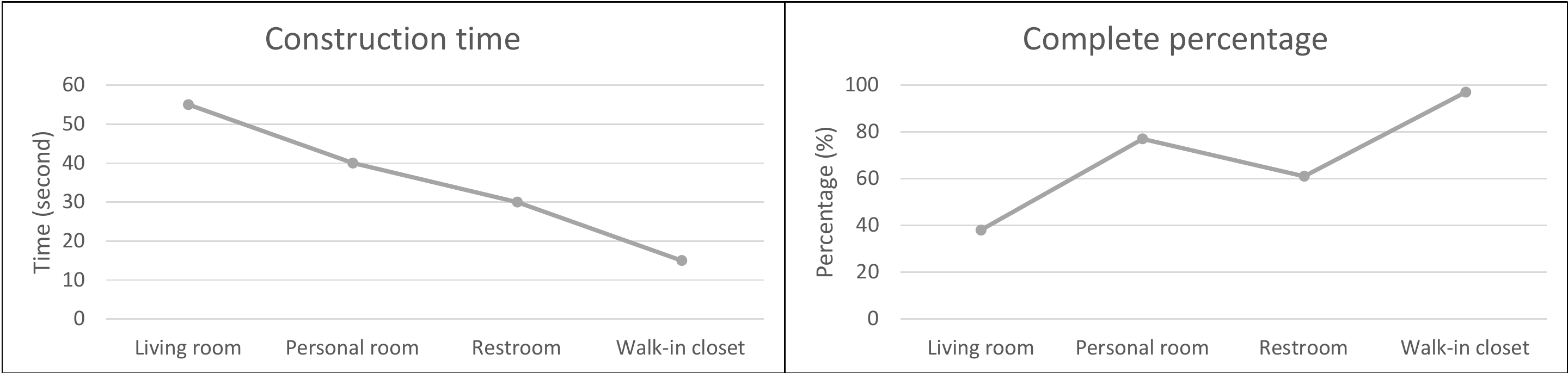}
\caption{Real-world rooms reconstruction time and complete percentage}
\label{fig:realComplete}
\vspace{-1em}
\end{figure}

In this experiment, we evaluate the real-world environment construction result in different room sizes and different illumination. Based on Figure \ref{fig:realComplete}, larger rooms need more time to generate a holographic room. Also, better illumination can affect construction speed and the completion percentage. From our experimental results, we can find a unique situation in the graph that the construction completion percentage of the biggest living room is not good enough. The reason is that the HoloLens 2 can only support about 50\% of living room hologram in the app simultaneously. Once the testers walk from one side to the other side, the device will automatically drop some hologram to reduce the load on the system, as shown in Figure \ref{fig:maxHologram}. In addition, one thing that users or developers should be aware of is that the irregular real-world corner could be challenging for HoloLens 2 to identify and generate, if the person stands far away. Therefore, users should face the front of the corner to help the device construct the corner hologram before using an app, an example shows in Figure \ref{fig:exIncomplete}.

\begin{figure}[htbp]
\centering
\includegraphics[width=8.4cm, keepaspectratio=true]{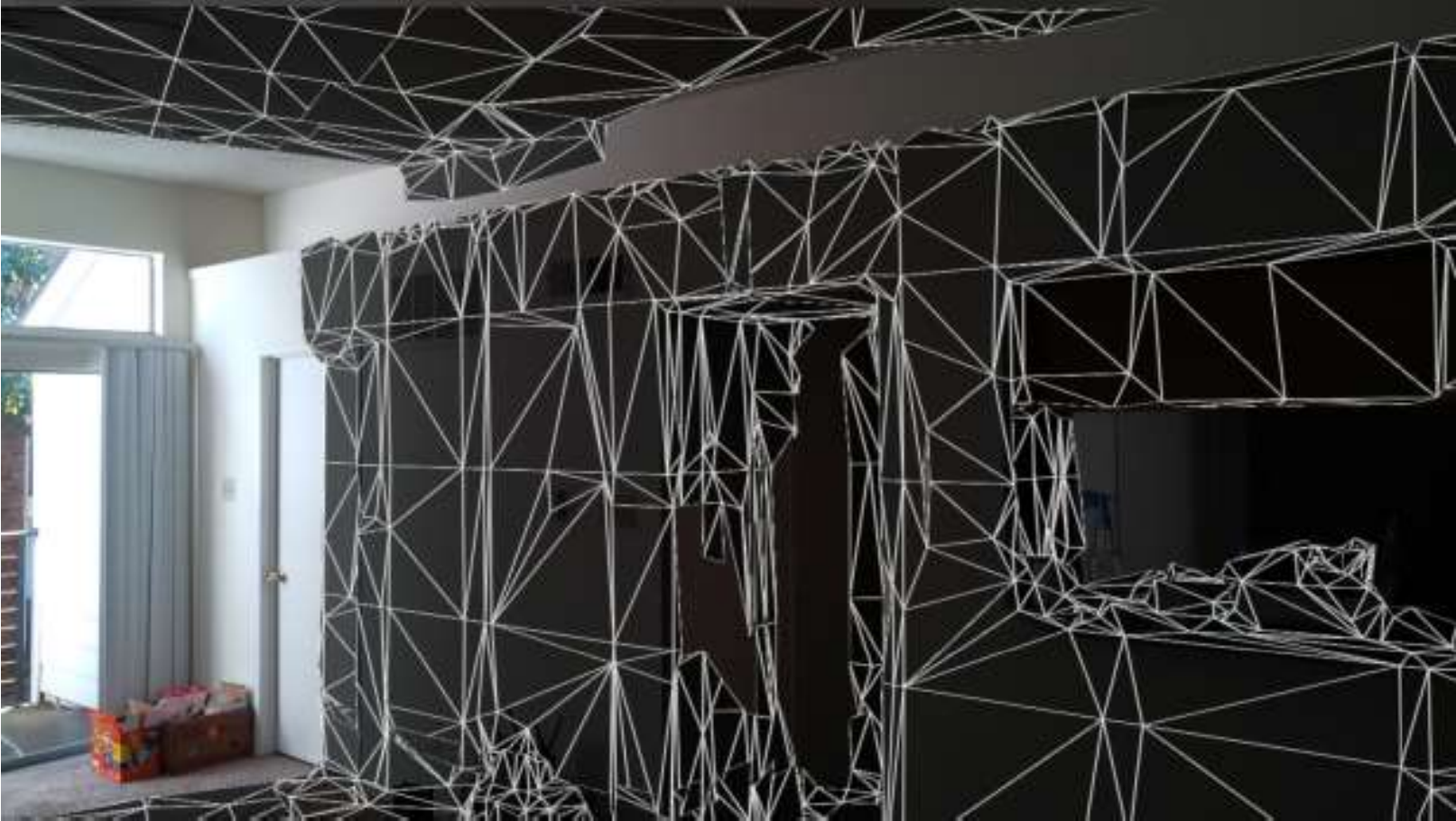}
\caption{An example of HoloLens 2's maximum hologram in the largest room in our experiment}
\label{fig:maxHologram}
\vspace{-1em}
\end{figure}

\begin{figure}[htbp]
\centering
\includegraphics[width=8.4cm, keepaspectratio=true]{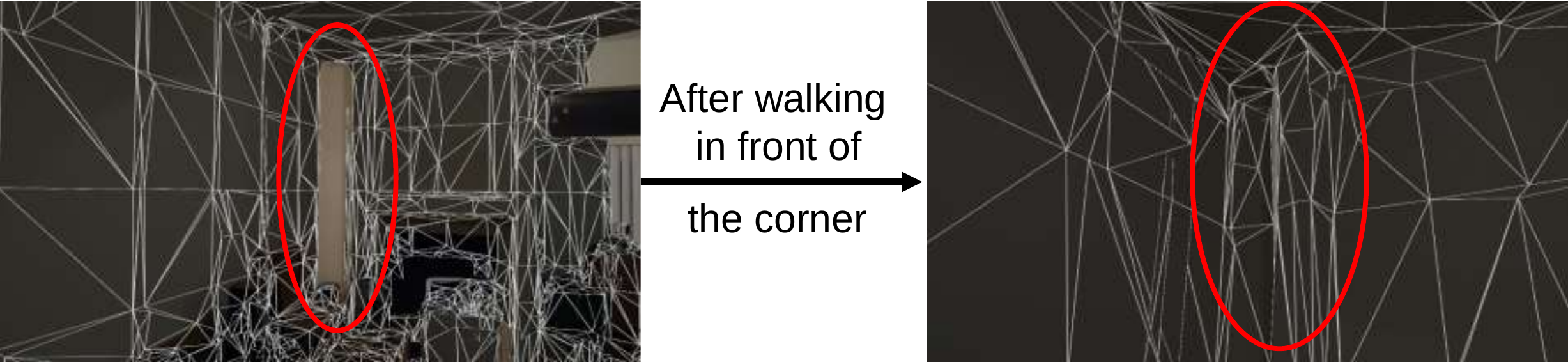}
\caption{An example of incomplete construction of a room's corner. The red circle on the left side indicate the incomplete construction part in the corner, and the red circle on the right side means the complete construction in the same corner}
\label{fig:exIncomplete}
\vspace{-1em}
\end{figure}

\end{enumerate}

\subsection{Speech Recognition}
\begin{figure}[htbp]
\centering
\includegraphics[width=8.4cm, keepaspectratio=true]{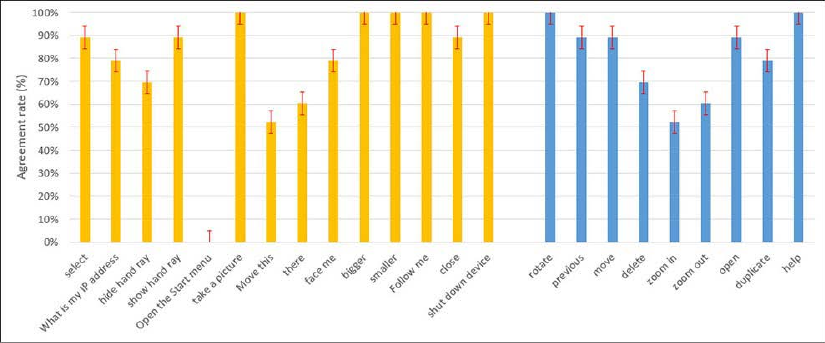}
\caption{Speech recognition agreement rate for the eighteen testers for the user-defined commands and system-defined commands}
\label{fig:speechAgree}
\vspace{-1em}
\end{figure}

In the speech recognition experiment result, we use the similar speech commands as the paper \cite{Liu2018}. Also, we use the same evaluation method, agreement rate, in our experiment result. The agreement rate \(A_{r}\) indicates the level of consensus among the participants for a specific referent \textit{r} and is defined as
\begin{equation}
\begin{aligned}
A_{r} = \sum_{P_{i}\in P_{r}}^{} \left ( \frac{\left | P_{i} \right |}{\left | P_{r} \right |} \right )^{2}\qquad
\end{aligned}
\end{equation}
where \(P_{r}\) is the set of operation commands for referent \textit{r} and \(P_{i}\) is a subset of \(P_{r}\). The speech recognition agreement rate result of HoloLens 2 is shown in Figure \ref{fig:speechAgree}, yellow and blue bars represent the agreement rates \(A_{r}\) for the system-defined commands and user-defined commands. 

Base on Figure \ref{fig:speechAgree}, the average agreement rates \(A_{r}\) for the system-defined commands and the user-defined commands are 79.12\% and 80.97\%. According to the HoloLens 1 speech recognition result \cite{Liu2018}, the average agreement rates \(A_{r}\) for the system-defined commands and the user-defined commands are 66.87\% and 74.47\%. Therefore, we can find a small improvement of the user-defined commands between HoloLens 2 and the previous device. On the other hand, for the system-defined commands, the speech recognition functionality of HoloLens 2 is much better than HoloLens 1; this is a significant improvement from the previous version to the current version. Besides, one thing worth mentioning is the command "Open the start menu," which is represented on the HoloLens 2 instruction document. However, none of our testers could successfully issue this command.

\subsection{Azure Spatial Anchor}
\begin{figure}[htbp]
\centering
\includegraphics[width=8.4cm, keepaspectratio=true]{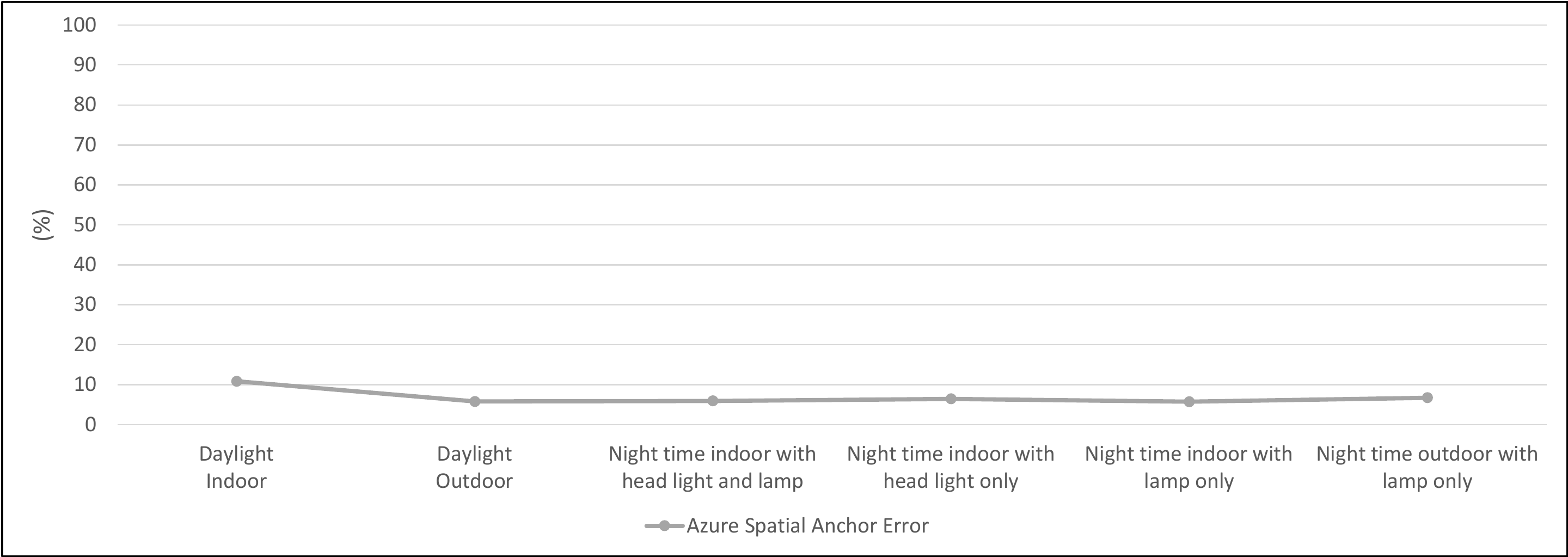} 
\caption{The Azure Spatial Anchor error distance percentage in six different environments}
\label{fig:AzureError}
\vspace{-1em}
\end{figure}

To calculate the error distance percentage between the created spatial anchor and the server downloaded spatial anchor, we use the blue and red lines show in Figure \ref{fig:Azure}b as a benchmark to calculate the ratio. Since the length and wide of the white table in real-world in each picture will be the same, we can compute the error distance by the following equation:
\begin{equation}
\begin{aligned}
{\rm Error Percentage}(\%) = \left ( \frac{\left | \frac{L_{c}^{s}}{L_{c}^{l}} - \frac{L_{d}^{s}}{L_{d}^{l}} \right |}{\frac{L_{c}^{s}}{L_{c}^{l}}} + \frac{\left | \frac{W_{c}^{s}}{W_{c}^{l}} - \frac{W_{d}^{s}}{W_{d}^{l}} \right |}{\frac{W_{c}^{s}}{W_{c}^{l}}}\right ) * 100
\end{aligned}
\end{equation}
where L and W indicate length and width, s and l present short line segments and long line segments, and c and d indicate the created and downloaded anchors in real-world. We ask the testers to do this test process three times in one environmental condition and do it in six different environments. The result shows in Figure \ref{fig:AzureError}. \par

We find that the error percentages are all very small, which shows the consistency from creating the Azure Spatial Anchor in HoloLens 2 to retrieve it back from the Azure server. In this case, we can easily and accurately use this feature to collaborate with others by creating spatial anchor apps in HoloLens 2, Android, or Apple devices.  

\subsection{Research Mode -- IMU sensors}
In the IMU sensors experiments, we use an error percentage to calculate the difference between the IMU sensors output result and actual movement:
\begin{equation}
\begin{aligned}
{\rm IMU Error} (\%) = \frac{1}{N}\sum_{i=1}^{N}\frac{|T_{N} - P_{N}|}{T_{N}}
\end{aligned}
\end{equation}
where N is the number of total tests, \(T_{N}\) indicates the true movement result of the test N and \(P_{N}\) expresses the device predicted result of the test N.
\begin{enumerate}
\item Accelerometer: \hspace{\fill}\\
For the first evaluation test of the Accelerometer, we ask the testers to walk in four different directions and different distances, which are straight, backward, right, and left with 3 or 4 meters, the testing environment setting shows in Figure \ref{fig:IMU}a. Then we compute the moving distance generated by the device and compare it with the actual moving distance with \textit{IMU Error}. The result shows in Figure \ref{fig:resultIMU}a. One thing to note is that the gravity acceleration will affect the Accelerometer result; once the time gets longer, the error will become larger. That is the reason why IMU sensors need to be used together to achieve complementary effects. Therefore, using the Accelerometer alone will make the output result higher than expected and cause unstable results. \par
In the second Accelerometer experiment, we ask the testers to tilt their body in two different directions three times to calculate the average error. One is to bow forward and tilt 90 degrees, altering their body along the x-axis and y-axis, an example shows in Figure \ref{fig:IMU}d. The other is to tilt 90 degrees to the right, which means moving their body along the X-axis and Z-axis. The result shows in Figure \ref{fig:resultIMU}b, the error percentage value here is small, which indicates that the Accelerometer performs well in the angle-changing situation.

\item Gyroscope: \hspace{\fill}\\
In the Gyroscope evaluation process, we ask the testers to rotate the HoloLens 2 device along the X, Y, and Z axes in three different ways. Then, we can calculate the error percentage by using \textit{IMU Error}. The result shows in Figure \ref{fig:resultIMU}c. The rotation error percentage is very low and these errors lie within operation errors. 

\begin{figure}[htbp]
\centering
\includegraphics[width=8.4cm, keepaspectratio=true]{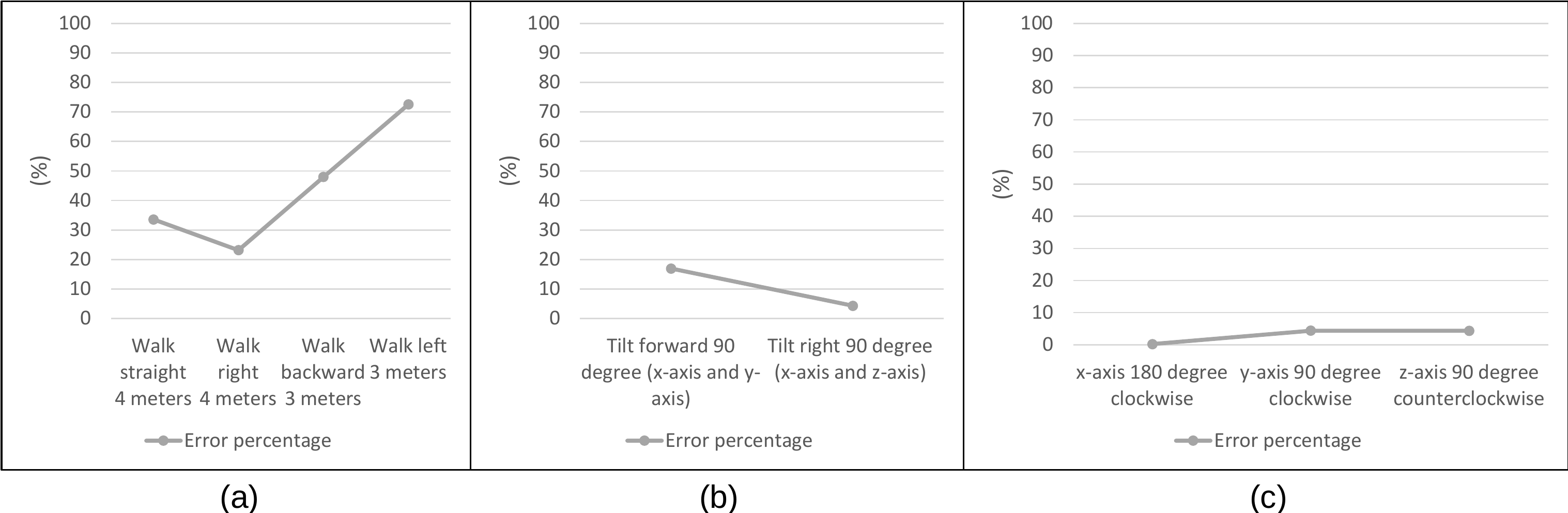}
\caption{The testing result of IMU sensors evaluation (a) Accelerometer moving distance error percentage (b) Accelerometer tilt angle error percentage (c) Gyroscope rotate error percentage}
\label{fig:resultIMU}
\vspace{-1em}
\end{figure}

\end{enumerate}


\section{Conclusion}
In this paper, we evaluated most of the important features in the HoloLens 2 \cite{hololensWeb} by building some related apps to show their capacities and limitations. Also, we explored the new version of research mode \cite{ungureanu2020hololens} in the current device and showed its usability. Then, we compared with the previous paper \cite{Liu2018} with similar experimental content to show the evolution between the two versions of HoloLens. These results can be used both as a benchmark and design criteria for MR applications. At the same time, these can be used as a reference and comparison target for other AR devices. Furthermore, users can use the evaluation results of each sensor of HoloLens 2 as a reference to conduct additional studies. \par
In the future, we plan to evaluate other features of HoloLens 2 and use the existing results for different research objectives.

\section{Declarations}
\textbf{Funding} No funding was received.\\
\textbf{Conflicts of interest} Not applicable.\\
\textbf{Availability of data and material} Currently unavailable.\\
\textbf{Code availability} Currently unavailable.

\bibliographystyle{spphys}
\bibliography{HoloLens_2_Technical_Evaluation_as_Mixed_Reality_Guide.bib}

%
%

\end{document}